%% file: tmi.tex
\documentclass[journal]{IEEEtran}
\usepackage{cite}
\usepackage{amsmath,amssymb,amsfonts}
\usepackage{algorithmic}
\usepackage{graphicx}
\usepackage{textcomp}

\usepackage{soul}
\usepackage{booktabs}
\usepackage{algorithm}
\usepackage{color,soul}
\usepackage{caption}

\usepackage{orcidlink}

\newcommand{\DM}{{\rm D}_{M}}
\newcommand{\forC}[1]{{\rm C}_{#1}}
\newcommand{\forPD}[1]{{\rm PD}_{{\rm C}_{#1}}}
\newcommand{\forPI}[3]{{\rm {#1}}_{{\rm C}_{#2}, {#3}}}
\newcommand{\forShortPI}[1]{{\rm \Pi}_{#1}}
\newcommand{\piatt}{PI-\textit{Att}}

\usepackage{tikz}
\newcommand\copyrighttext{%
  \footnotesize This work has been submitted to the IEEE for possible publication. Copyright may be transferred without notice, after which this version may no longer be accessible.}
\newcommand\copyrightnotice{%
\begin{tikzpicture}[remember picture,overlay]
\node[anchor=south,yshift=10pt] at (current page.south) {\parbox{\dimexpr\textwidth-\fboxsep-\fboxrule\relax}\copyrighttext};
\end{tikzpicture}%
}

\def\BibTeX{{\rm B\kern-.05em{\sc i\kern-.025em b}\kern-.08em
    T\kern-.1667em\lower.7ex\hbox{E}\kern-.125emX}}
\begin{document}
\title{\piatt: Topology Attention for Segmentation Networks through Adaptive Persistence Image Representation}
\author{Mehmet Bahadir Erden, Sinan Unver, Ilke Ali Gurses, Rustu Turkay, Cigdem Gunduz-Demir
\thanks{This study was supported by Scientific and Technological Research Council of Turkey (TUBITAK) under the Grant Number 120E497. The authors thank to TUBITAK for their supports.}
\thanks{M.B. Erden is with the Department of Computer Engineering and KUIS AI Center, Koc University, 34450 Istanbul, Turkey (e-mail: merden22@ku.edu.tr).}
\thanks{S. Unver is with the Department of Mathematics, Koc University, 34450 Istanbul, Turkey (e-mail: sunver@ku.edu.tr).}
\thanks{I A. Gurses is with the School of Medicine, Koc University, 34010 Istanbul, Turkey (e-mail: igurses@ku.edu.tr).}
\thanks{R. Turkay is with the Department of Radiology, School of Medicine, Haseki SUAM, Medical Sciences University, 34265 Istanbul, Turkey (e-mail: rustu.turkay@sbu.edu.tr).}
\thanks{C. Gunduz-Demir is with the Department of Computer Engineering, School of Medicine, and KUIS AI Center, Koc University, 34450 Istanbul, Turkey (e-mail: cgunduz@ku.edu.tr).}}

\maketitle

\copyrightnotice

\begin{abstract}
Segmenting multiple objects (e.g., organs) in medical images often requires an understanding of their topology, which simultaneously quantifies the shape of the objects and their positions relative to each other. This understanding is important for segmentation networks to generalize better with limited training data, which is common in medical image analysis. However, many popular networks were trained to optimize only pixel-wise performance, ignoring the topological correctness of the segmentation. In this paper, we introduce a new topology-aware loss function, which we call \piatt, that explicitly forces the network to minimize the topological dissimilarity between the ground truth and prediction maps. We quantify the topology of each map by the persistence image representation, for the first time in the context of a segmentation network loss. Besides, we propose a new mechanism to adaptively calculate the persistence image at the end of each epoch based on the network's performance. This adaptive calculation enables the network to learn topology outline in the first epochs, and then topology details towards the end of training. The effectiveness of the proposed \piatt~loss is demonstrated on two different datasets for aorta and great vessel segmentation in computed tomography images.
\end{abstract}


\input{parts/1_intro}

\input{parts/2_related_work}
\input{parts/3_method}
\input{parts/4_experiments}

\input{parts/5_results_discussion}
\input{parts/6_conclusion}

\bibliographystyle{IEEEtran} 
\bibliography{tmi}

\end{document}

%% file: parts/1_intro.tex
\section{Introduction}
\label{intro}

\begin{figure}[b]
\centering
\includegraphics[width=5.8cm]{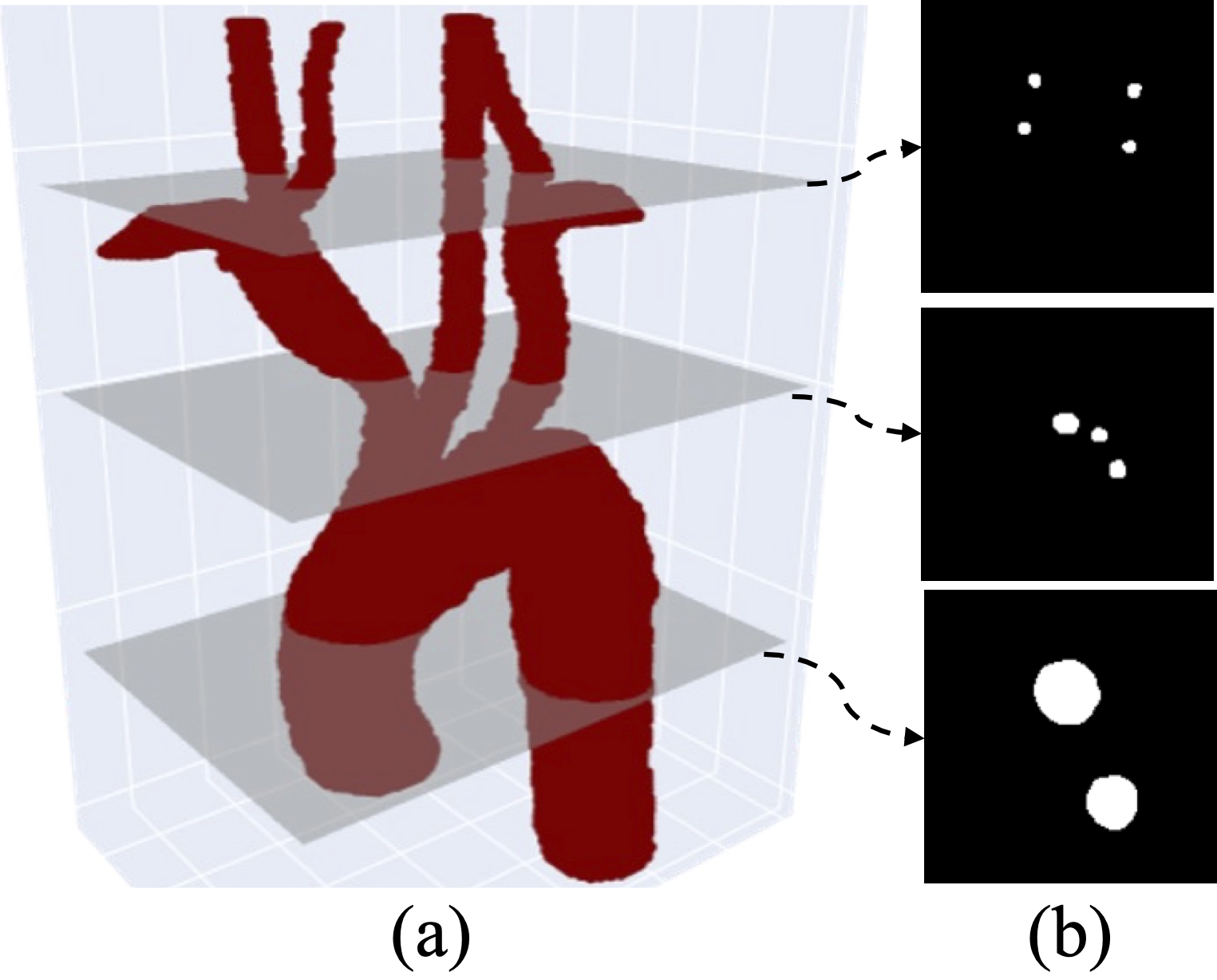} 
\captionsetup{font = small}
\caption{(a) 3D visualization of the aortic arch and great vessels (large arteries and veins). (b) Segmentation maps of axial slice examples. Vessels in an axial slice are not randomly distributed in the body but found in a particular topology, which differs according to position of the axial slice with respect to the heart.} 
\label{fig:aorta}
\end{figure}

\IEEEPARstart{E}{ncoder-decoder} networks are commonly used for the segmentation of organs or anatomical structures in medical images. Although they share similar architectures with those designed to segment natural images, there are two main distinctive aspects of medical image segmentation. On one hand, since it is quite challenging to obtain pixel-level annotations for medical images, the networks must be trained with very limited data, which necessitates more effective uses of regularization techniques in training. On the other hand, due to the human anatomy, there is prior knowledge regarding the organs/structures in the body even though there exist anatomical variations in the organs/structures as well as noise and artifacts in the images. One may use this prior knowledge in the network design as a regularization technique. To this end, previous studies commonly calculated shape descriptors on the organs/structures and forced the network to minimize the shape inconsistency between the ground truth and prediction. This was typically achieved by multi-task networks with an additional task of learning these descriptors~\cite{wang2020deep, murugesan2019psi} or losses with an extra term that penalized the shape inconsistency~\cite{al2018shape, wong2021persistenthomology}. When there is one organ/structure to be segmented, forcing the network to learn its shape may help better generalizations. However, when there are multiple organs or anatomical structures to be segmented and when they are found in the body in an expected topology (Fig.~\ref{fig:aorta}), it is more beneficial to force the network to preserve this topology along with learning the shape correctness.

\begin{figure*}[!htb]
\vspace{-0.25cm}
\centering
\includegraphics[height=6.4cm]{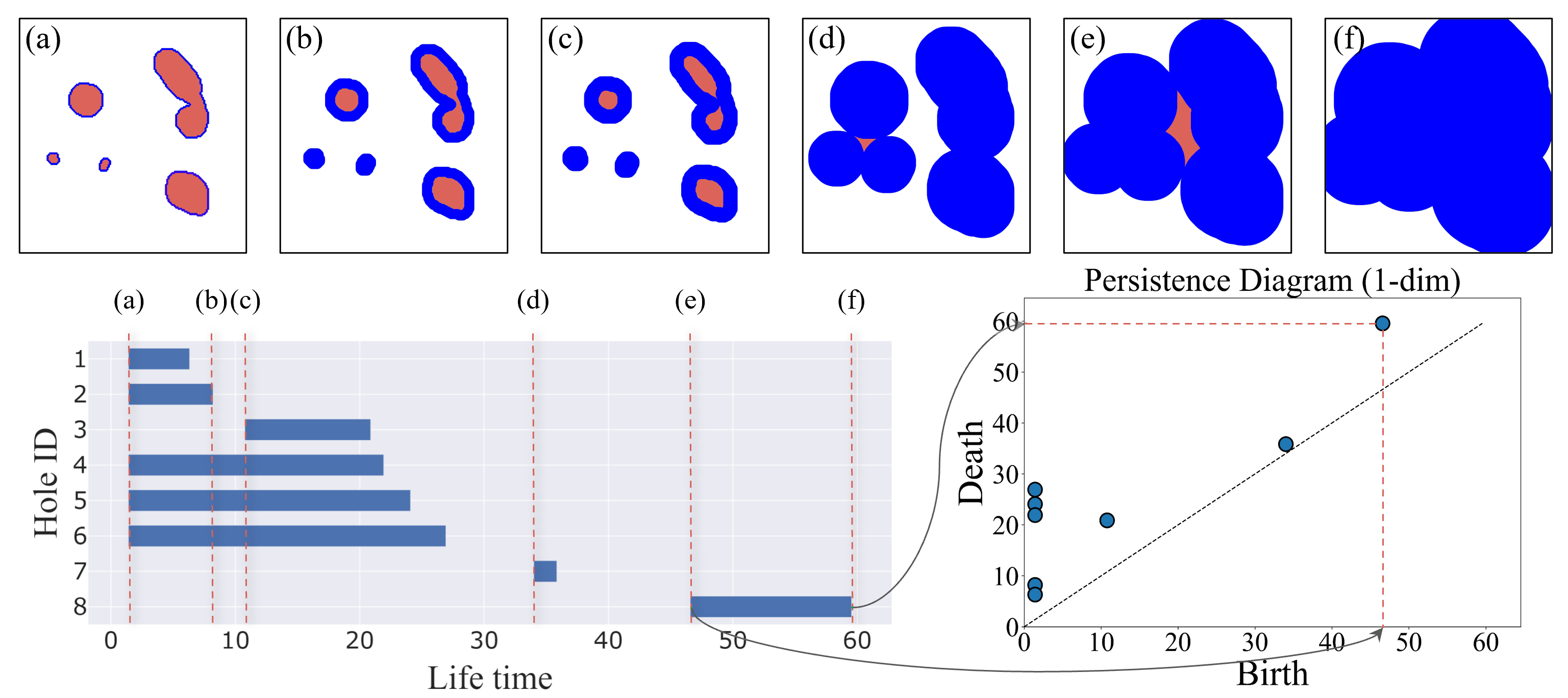}
\captionsetup{font = small}
\vspace{-0.1cm}
\caption{Illustration of 1-dimensional persistent homology and the process of constructing a persistence diagram using a distance filtration on contour points. This filtration was selected only for illustration, as it is easier to see how persistent homology simultaneously models shape and geometry. Our model will use another filtration also defined on the contours but associated to a density function as it gives better topological summary. The bottom indicates the filtration process with the barcodes and depicts the resulting persistence diagram. The top shows the evolved contour points (blue) and holes (red) at six time indices, from (a) to (f) and with dashed lines. (a) At time 0, five holes are born inside the initial contours of the objects. (b) Two holes die as their corresponding objects are smaller than the others. (c) Another hole is born inside the object at the upper-right corner. The shape of this object (figure-eight-like shape) is different from the others. (d) All holes inside the objects die but another hole between three objects is born. Since these objects are closer to each other than the remaining two, a hole between them is born before the final hole illustrated at time (e). The barcodes of the last two holes are associated with the objects' geometry. (f) All holes die at the end.}
\vspace{-0.3cm}
\label{fig:persistent_homology}
\end{figure*}

Persistent homology emerges as a mathematical tool in computational topology. With an appropriate selection of its filtration function, it can simultaneously model the shape of an individual object and the geometry of the total configuration of all objects. 1-dimensional persistent homology, which our \piatt~loss will use, models how holes evolve over the filtration process by representing the timespan of a hole during this process with a barcode. In the equivalent persistence diagram representation, each barcode is represented by a point whose coordinates are the time indices when the corresponding hole is born and when it dies. Intuitively, points (or barcodes) associated with the holes inside the contours of a single object quantifies its shape and size whereas points associated with the holes formed among the contours of different objects quantifies the geometry of the total configuration (Fig.~\ref{fig:persistent_homology}).

\begin{figure*}[t]
\vspace{-0.22cm}
\centering
\includegraphics[width=0.74\textwidth]{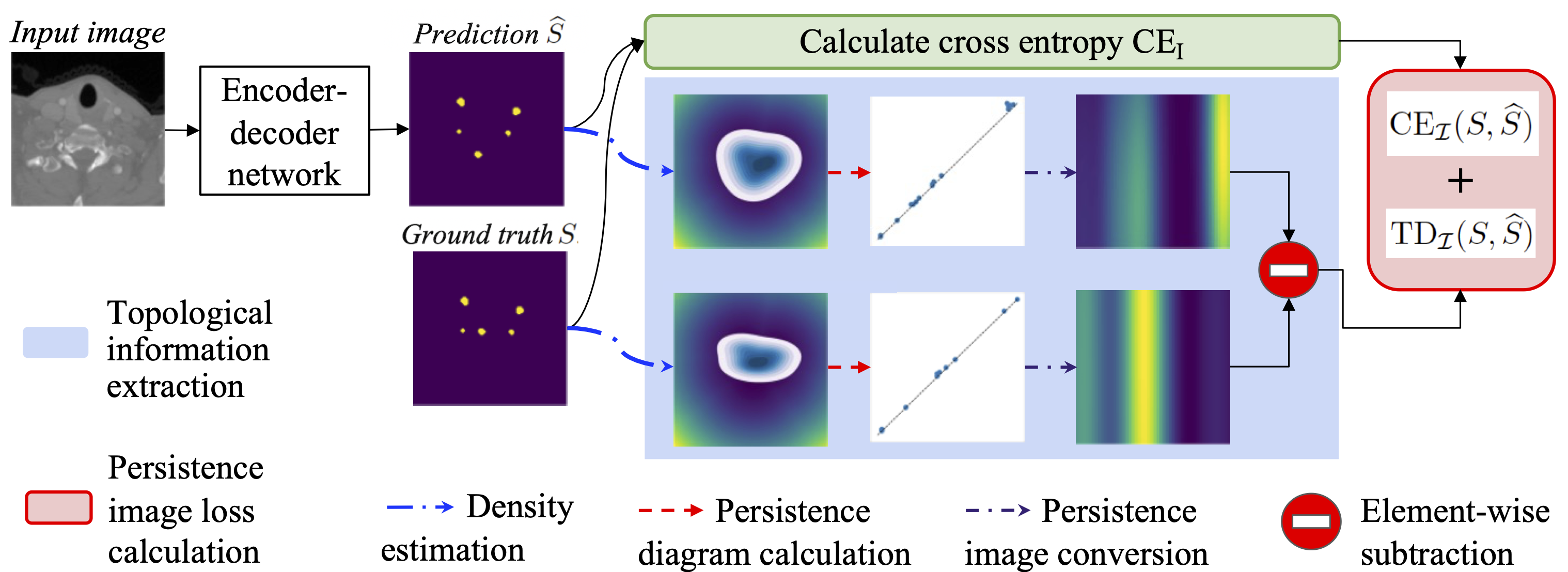}
\captionsetup{font = small}
\vspace{0.2cm}
\caption{Schematic overview of the proposed \piatt~loss.} 
\label{fig:overview}
\vspace{-0.3cm}
\end{figure*}

Previous studies integrated persistent homology into the design of a segmentation network by defining custom losses that penalized dissimilarities between the persistence diagrams~\cite{wong2021persistenthomology, hu2019topologypreservingdeep} or persistence barcodes~\cite{clough2022topologicalloss} of ground truth and prediction maps. The dissimilarity of two diagrams was calculated by finding matches between the points in the diagrams, minimizing the cost over all matching, and using the distances between the matched points. A point in one persistence diagram is matched either with an unmatched point in the other diagram or with its closest point on the diagonal, if there exists no matching point in the other diagram. Two common dissimilarity metrics are the bottleneck distance, the maximum distance between the matched points, and the Wasserstein distance, the sum of the powers of distances between all matched points.

This dissimilarity calculation has drawbacks: First, the point matches for the persistence diagrams of the ground truth and prediction maps may be unsteady before the network starts to converge. Besides, even an extra point associated with a false positive hole in the prediction map or a missing point associated with a false negative may lead to drastic changes in the point matches. This causes volatility in the calculated distances, and thus in the loss, which may prevent steady learning. Although this problem was also mentioned in the literature~\cite{oner2022persistenthomology}, the filtration methods proposed for this purpose still had drawbacks such as limited direction consideration in filtering because of the computational limits. \textit{Our work addresses this problem by using the persistence image representation\cite{adams2017persistence}. In particular, instead of matching the points of the two diagrams, it proposes to calculate the persistence image of each diagram separately and then to use the difference between these two persistence images as the topological dissimilarity term in a loss function.} Second, since the bottleneck distance only models the phenomenon linked to one match with the maximum distance, it may be insufficient to model the shape and topology of multiple objects. The same problem arises when only the most persistent (longest) barcode is considered. Although the Wasserstein distance mitigates the problem, it may be negatively affected by treating all matches the same. The points far from the diagonal are linked to holes that live longer, so the matches involving these points typically model the dissimilarity between the general outlines of the shape and topology of the objects in the ground truth and prediction. On the contrary, the points close to the diagonal live shorter and are often associated with the details. One may want to penalize the distance for the former type of matches in earlier epochs, as learning the general shape and topology outline is more important when the network starts to learn, and penalize the distance for the latter type in later epochs, as the details become more important towards the end of training. Previous studies did not handle such kind of adaptation. \textit{Our work addresses this problem by introducing an adaptive scheduler mechanism for persistence image calculation based on the network's performance in each training step.} The schematic overview of the proposed approach is given in Fig.~\ref{fig:overview}. 

The contributions of this paper are summarized as follows:
\begin{itemize}
    \item This is the first proposal of using the persistence image representation in the context of a segmentation network loss. We propose to define a custom loss, \piatt, with the topological dissimilarity term calculated on the persistence image representations of ground truth and prediction maps. This proposal eliminates the problematic point matching occurred when the bottleneck or Wasserstein distance is used, as suggested by the previous studies.
    
    \item We introduce an adaptive scheduler mechanism to dynamically calculate the persistence images at the end of each training step, based on the network performance. This mechanism helps learn topology outline first when learning starts, and focus on details later as time passes. Such adaptive persistence image calculation has not been proposed by the previous studies. 
    
\end{itemize}

 We tested our proposed model on two datasets of computed tomography (CT) images of the aorta and great vessels. Our experiments demonstrated that this model offered a promising tool to regularize the training of a segmentation network even with limited data, improving the results of its counterparts.

%% file: parts/2_related_work.tex
\section{Related Work}

\underline{Persistent homology:} As one of the most important mathematical tools in computational topology, it was applied to many domains. Medical domain examples include generating persistent homology profiles from patients’ images~\cite{qaiser2016persistenthomology}, enhancing classifier performance using topological features extracted on such profiles~\cite{qaiser2018fastaccurate,singh2014topologicaldescriptors}, and predicting survival rates based on their statistics~\cite{somasundaram2021persistenthomology}. 

\underline{Topology and neural networks:} Topology was used to measure the complexity of networks~\cite{rieck2019neuralpersistence} and to gain insight into how they function~\cite{gabrielsson2018expositioninterpretation}. In generative models, the construction loss was advanced by topology comparison in the input/output or latent/feature space~\cite{gabrielsson2020topologylayer, hofer2019connectivityoptimizedrepresentation, moor2020topologicalautoencoders, charlier2019phomgempersistent, wang2020topogantopologyaware}. In classification networks, persistent homology enabled to embed a topological penalty as a regularizer~\cite{chen2019topologicalregularizer} and extract topological features~\cite{khramtsova2022rethinkingpersistent} for performance improvement. Topology was incorporated into segmentation networks by mostly defining a custom loss. The study in~\cite{hu2019topologypreservingdeep} addressed having minor structural discrepancies (e.g., broken connection in an object) in a prediction map. It defined a topological loss by comparing the Betti numbers of the prediction likelihood and the corresponding ground truth, and used the Wasserstein distance to calculate the difference between the two persistence diagrams. Other studies also defined their losses using the Wasserstein distance between the persistence diagrams of the ground truth and prediction maps; these studies differed in their distance calculations~\cite{haft-javaherian2020topologicalencoding, ozcelik2023topologyawareloss,demir2023topologyawarefocal}. 

The study in~\cite{clough2022topologicalloss} presented an alternative approach that used the zeroth and first Betti numbers as topological priors on connected components and holes, respectively. It forced the network to predict maps producing the same Betti numbers with the desired values in the ground truth through barcode diagrams. It was also extended to multi-class segmentation~\cite{byrne2023persistenthomologybased}. Lastly, persistent homology was employed for segmentation, using the discrete Morse theory, to prune undesired parts of the Morse structure~\cite{hu2021topologyawaresegmentation}. Nevertheless, all these models directly used the persistence diagrams or barcode diagrams; none of them used the persistence image representation. 

\underline{Persistence image representation:} There exist only a few studies that used the persistence image representation. Different than our proposal, they used it as a feature extractor or in an optimization process, but not to define a custom segmentation loss. In~\cite{samani2021visualobject}, the QR-pivoting technique was applied on persistence image representations to solve sparse topological information, and its output was fed as an input to an object recognition network. In~\cite{zhao2019learningmetrics}, weighted persistence image kernels were used in the optimization phase with the metric learning objective to advance the graph classification performance. In~\cite{chen2021zgcnetstime}, the persistence image calculated with a zigzag filtration was utilized as a topological feature to improve performance in time series forecasting. In~\cite{wong2021persistenthomology}, the persistence image representation was used to generate flattened topological feature vectors, which were then concatenated in feature maps of a network. Although this previous study designed a segmentation network that used a topological loss, it calculated this loss based on the Wasserstein distance between the persistence diagrams but did not use the persistence images in any step of the loss definition. Besides persistence images, the persistence landscape representation was used as a feature extractor in a network to enhance classification performance~\cite{khramtsova2022rethinkingpersistent, kim2021pllayefficient, liu2016applyingtopological}. However, it is different than the persistence image representation as well as it is not used to define a loss function. Thus, all these studies were different than our proposal. 

%% file: parts/3_method.tex
\section{Method}
\label{sec:method}

\subsection{Persistence Image Representation}
Let $M$ be a segmentation map (either a ground truth or a prediction map) and $\forC{M}$ be its contour points. This work quantifies the map $M$ by calculating the persistence image representation on the corresponding contour points $\forC{M}$. The pseudocode is in Algorithm~\ref{algo:pers-img} and details are provided below.
\begin{algorithm}[t]
\caption{{\sc PersistenceImageCalculation}}\label{algo:pers-img}
\footnotesize
\textbf{Input:} map $M$, bandwidth $B$ of a density estimator kernel, variance $\sigma^2$ of the Gaussian distribution, and parameter $\gamma$ of the weighting function. \\
\textbf{Output:} normalized persistence image $\forShortPI{M}$ of the map $M$. 
\vspace{-0.1cm}
\begin{algorithmic}[1]
\STATE $\forC{M} =$ {\sc FindObjectContours}$(M)$
\STATE $\forPD{M} =$ {\sc CubicalComplexFiltration}$(\forC{M}, B)$
\STATE $\forShortPI{M}$ = 0
\FOR {\textbf{each} point $(b,d) \in \forPD{M}$}
\STATE $(x, y) = (b, d-b)$
\STATE $g_{(x,y)} =$ {\sc ApplyGaussianFiltering}$((x,y), \sigma^2) $
\STATE $\forShortPI{M} = \forShortPI{M} + \omega(y, \gamma) \cdot g_{(x,y)}$
\ENDFOR
\STATE $\forShortPI{M} =$ {\sc z-Normalization}$(\forShortPI{M})$
\end{algorithmic}
\end{algorithm}

\subsubsection{Persistent Homology} To quantify the topology of data, it is natural to consider the data as a topological space and study its homology groups. A $d$-degree homology group of a topological space is a linear invariant which contains information of how many independent $d$-dimensional holes there are in the space. This naive approach has major problems: First, real-world data are inherently noisy but homological information is very rigid such that it could change even if a single point is removed from the topological space. Second, the real-world data are discrete and discrete spaces have uninteresting homology groups. This can be solved by making spaces under consideration continuous by replacing the space with the space of points within $\varepsilon$ distance of the set, for some $\varepsilon >0.$ This idea results in obtaining interesting homology groups but creates another problem that information depends on the auxiliary choice of $\varepsilon.$ Carlsson~\cite{carlsson2009topologydata} addressed these problems through his discovery of persistent homology. In our context, this can be thought of as considering all  $\varepsilon$ at the same time. 

If one starts with a topological space $X$ together with a filtration $\{F_{i}\}_{i >0},$ i.e., a subspace $F_{i}$ of $X$ for each $i>0$ such that  $F_{i} \subseteq F_{j} ,$ for $i \leq j,$ and $\cup _{i>0}F_{i} =X,$ then {\it persistence homology} associates to this data a {\it persistence diagram}. A persistence diagram is a finite collection of points $(b,d)$ with $0\leq b< d ,$ where $b$ and $d$ respectively denote the birth and death times of a hole throughout the filtration. Fig.~\ref{fig:persistent_homology} illustrates 1-dimensional persistent homology and the process of constructing a persistence diagram when the filtration is defined using a distance function. (In Fig.~\ref{fig:persistent_homology}, the distance function is selected for illustration as it is easier to understand the process. Our model will use another filtration that gives a better topological summary of an image with discrete pixels.)

\begin{figure*}[t]
\vspace{-0.2cm}
\centering
\includegraphics[height=7.2cm]
{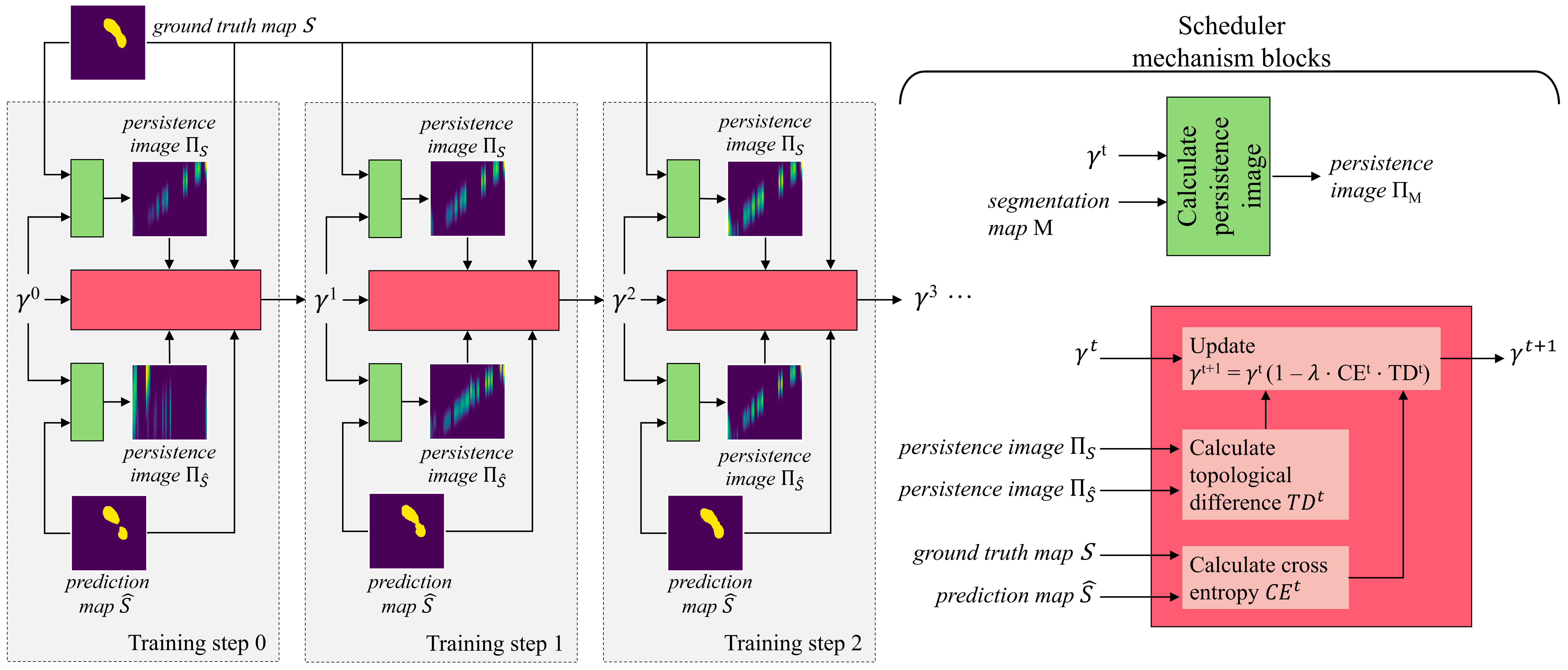}
\captionsetup{font = small}
\caption{Illustration of the proposed adaptive scheduler mechanism for three different steps of network training. At the top, the existing ground truth of an example image and its persistence images calculated in different steps are given. At the bottom, the segmentation map predicted in a given step and its persistence image are given. All persistence images are calculated using $\gamma^t$ of the corresponding step. The scheduler blocks used in this mechanism are shown as green and red boxes, and their operations are depicted on the right.}
\label{fig:gamma_update_mech}
\vspace{-0.1cm}
\end{figure*}

\subsubsection{Persistence Diagram Calculation.} Our model uses the persistence diagram of a cubical complex together with a filtration associated to a density function. Let $\DM$ denote the set of all pixels in a grid corresponding to a segmentation map $M$ and $\forC{M} \subseteq \DM$ denote the segmentation contours in $M$. Associated to $\forC{M},$ we form the kernel density estimator using a Gaussian kernel with a bandwidth $B$.

This density estimator defines a function $f_{\forC{M}}:\DM \to (0,1),$ depending on $\forC{M}.$ Letting $g_{\forC{M}}:= -\log \circ f_{\forC{M}}$ and defining $F_{i}$ to be the inverse image $g_{\forC{M}}^{-1} ((0,i])$ of $(0,i]$ under $g_{\forC{M}},$ for every $i \in (0,\infty),$ we obtain an increasing filtration $\{ F_{i} \}_{i>0}$ of $\DM.$ When $i$ is small, the level-$i$ of the filtration $F_{i}$ only admits points in $\DM$ whose vicinity contains a large number of points from $\forC{M}.$ As $i$ increases, $F_{i}$ admits those whose vicinity contains less points from $\forC{M}$ and eventually as $i$ becomes very large $F_{i}$ contains all points of $\DM.$ We consider 1-dimensional persistent homology of this filtration and represent it with the persistence diagram $\forPD{M}.$ 1-dimensional persistent homology is related with the holes inside the contours of individual objects, which especially quantify the shape of the aorta, but also with the holes formed among the contours of different objects, quantifying the geometry of the total vessel configuration (Fig.~\ref{fig:persistent_homology}).

\subsubsection{Persistence Image Calculation}. The persistence diagram $\forPD{M} \subseteq \{(b,d)~|~ 0 \leq b <d \}$ is a finite collection of points, each of which is a pair indicating the birth and death times of a hole. The persistence image $\forPI{\Pi}{M}{f}$ is obtained from this persistence diagram by first applying a linear transformation $LT$ to the points $(b,d) \in \forPD{M}$ and then taking a weighted summation of the smoothed points in the new space~\cite{adams2017persistence}. Its pseudocode is given in Lines 3-7 of Algorithm~\ref{algo:pers-img}.

Let $LT:\mathbb{R}^{2} \to \mathbb{R}^{2}$ be the linear transformation given by $LT(b,d):=(b,d-b)$ to $\forPD{M}$ to obtain a finite set in the first quadrant of the Cartesian plane. In this new space, a point $u$ will have the coordinates of $(x, y)$ corresponding to the birth time $b$ of a hole and its lifetime defined as $d-b$, respectively. To  each point  $u \in  LT({\rm PD}_{C_M}),$ we attach a function  $g_u$, which is the normalized and symmetric Gaussian distribution with  mean $u$ and fixed variance $\sigma^2 = 1.0.$ To calculate the persistence image $\forPI{\Pi}{M}{f}$, one needs to fix a weighting function $f: \mathbb{R}^{2} \to \mathbb{R},$ which is non-negative, continuous, piecewise differentiable, and 0 on the $x$-axis. Fixing an $f,$ we define $\forPI{\rho}{M}{f} =\sum_{u \in LT(\forPD{M})}f(u)~g_u.$ To convert this to a function $\forPI{R}{M}{f},$ defined on a set of pixels, we let the value of $\forPI{R}{M}{f}$ on the pixel $p$ to be the integral $\int_{p} \forPI{\rho}{M}{f}~dx\wedge dy$ ~of~ $\forPI{\rho}{M}{f}$ on the region defined by $p.$ The values of $\forPI{R}{M}{f}$ give pixel intensities, which are then turned into an image called the persistence image $\forPI{\Pi}{M}{f}.$ Note that we use the persistence image on a fixed grid of pixels and z-normalize it to eliminate the problem of having scale differences between the persistence images of the ground truth and prediction maps as well as those belonging to different training images.

The persistence image will depend on the choice of the weighting function $f$. This work proposes to define it as a function $\omega(y, \gamma)$ 
of the lifetime $y$ of a hole and a parameter $\gamma$, which is dynamically scheduled at the end of each training step based on the network's performance. This function is detailed when we discuss the proposed adaptive scheduler mechanism in Sec.~\ref{sec:adaptive_mech}. But intuitively, one can think that this function gives more importance to holes with longer lifetimes in the first epochs of training, and reduces this importance by decreasing the value of the parameter $\gamma$, which makes all holes as important as the ones that live longer. Note that holes with longer and shorter lifetimes correspond to topology outline and its details, respectively. Thus, this weighting function allows to learn topology outline first, and then focus on also learning topology details towards the end of training. 

To increase the readability of the remaining equations, thereafter, we will refer to $\forPI{\Pi}{M}{f}$ ~as~ $\forShortPI{M}$ assuming that the persistence image is always calculated on the contours of the map $M$ by using the weighting function $\omega(y, \gamma)$.

\subsection{Topology-Aware Loss Definition}
\label{sec:persistence_img_loss}

For each image $\mathcal{I}$, the difference in the persistence images of its ground truth $S$ and the predicted segmentation map $\widehat{S}$ is used as the topological awareness term in the proposed loss \piatt. With $p$ denoting the pixels in these persistence images, the topological dissimilarity for the image $\mathcal{I}$ is defined as
\begin{equation}
\label{eq:topological_difference}
{\rm TD}_\mathcal{I}(S,\widehat{S})=\frac{1}{|\forShortPI{S}|} \sum_{~p \in \forShortPI{S}} |\forShortPI{S}(p) - \forShortPI{\widehat{S}}(p)|.
\end{equation}
This topological dissimilarity is considered as an attention term in updating the network weights and used with standard cross entropy ${\rm CE}_\mathcal{I}(S,\widehat{S})$ in a joint loss function given as
\begin{equation}
\mathcal{L}_\mathcal{I}(S,\widehat{S})= {\rm CE}_\mathcal{I}(S,\widehat{S}) + \beta \cdot {\rm TD}_\mathcal{I}(S,\widehat{S}),
\end{equation}
where $\beta = 0.05$ was empirically set. This definition enables a network to learn with cross-entropy while simultaneously correcting the topology of problematic segmentation maps. 

\subsection{Adaptive Scheduler Mechanism}
\label{sec:adaptive_mech}

A point in the persistence diagram affects the persistence image representation according to its position in the diagram, i.e., the birth and death times of the hole it corresponds to, and the weighting function. For a point $(x,y)$ in the new space $LT$, which is the linear transformation of the points in the persistence diagram, the weighting function $\omega$ is defined as 
\begin{equation}
\label{eqn:omega}
\omega(y, \gamma) = 
\begin{cases} 
y^\gamma & \text{if } \gamma \geq 1 \\
y & \text{if } \gamma < 1
\end{cases}
\end{equation}
where $y$ represents the lifetime of the hole (i.e., the vertical distance from the point to the diagonal) and $\gamma$ is the factor determining how strong pixels associated with the hole are represented in the persistence image. Points far away from the diagonal of the persistence diagram, i.e., holes with longer lives, typically correspond to topology outline and the ones close to the diagonal, i.e., holes with shorter lives, are usually associated with topology details. We propose to give more emphasis on the points corresponding to topology outline in the first epochs by initializing $\gamma^0 = 2$, and increase the importance of learning topology details towards the end of training by decreasing $\gamma$ using the proposed scheduler mechanism. This scheduler calculates $\gamma^{t+1}$ for the next training step $(t+1)$ as
\begin{equation}
\label{eqn:scheduler}
\gamma^{t+1}= \gamma^{t} \cdot (1 -  \lambda \cdot {\rm CE}^{t} \cdot {\rm TD}^{t})
\end{equation}
using $\gamma^{t}$ of the current training step, the overall cross entropy ${\rm CE}^{t} = \sum_{I \in {\mathcal T}} {\rm CE}^{t}_\mathcal{I}$, and the topology dissimilarity ${\rm TD}^{t} = \sum_{I \in {\mathcal T}} {\rm TD}^{t}_\mathcal{I}$ calculated on the ground truths and segmentation maps predicted for images in the training set ${\mathcal T}$ at the end of the current training step $t$.

We set $\lambda = 0.0005$, which leads to gradual decrease in $\gamma$ during training. For an example image, Fig.~\ref{fig:gamma_update_mech} visualizes the ground truth and segmentation maps predicted in three different epochs, and their persistence images calculated with $\gamma$ of the corresponding epoch. Since $\gamma$ gradually decreases in later epochs, the persistence diagram of the ground truth also changes, reflecting the fact that the network may emphasize different levels of topological details in different steps, even though the ground truth map is always the same. Note that the importance of the weighting function, in terms of flexibility in representing persistence diagram points and also contributing to the emphasis on different topological characteristics, was has already been pointed out in the previous studies~\cite{adams2017persistence,zhao2019learningmetrics,kusano2018kernelmethod}. However, none of them utilized the persistence image representation in a segmentation network or dynamically updated the weight parameter based on the segmentation performance.

%% file: parts/4_experiments.tex
\section{Experiments}

\subsection{Datasets}
We tested our loss on two datasets of CT images for the segmentation of aorta and great vessels. This application was chosen since anatomical variations of vessels may have clinical impacts (e.g., estimating complications during insertion of carotid artery stents or salivary bypass tubes), and their segmentation will be the first step to model and understand these variations. Additionally, for this application, due to the human anatomy, there exist expected shapes especially for the aorta, and an expected geometry of the total configuration of the aorta and great vessels. Thus, it provides an ideal showcase for incorporating topological awareness into the model.

The first is the in-house dataset that contains 2164 CT images of 24 subjects. We randomly split the subjects into three, to perform three-fold cross validation, and used all CT images of the same subject in the same fold. The data collection was conducted in accordance with the tenets of the Declaration of Helsinki and was approved by Koc University Institutional Review Board (protocol number: 2022.161.IRB1.06). We used this dataset for thorough analysis of our proposal. We then showed its applicability on the publicly available AVT dataset~\cite{radl2022avtmulticenter} examples taken from KiTS19 Grand Challenge~\cite{heller2020kits19challenge, heller2021stateart}. We also performed three-fold cross-validation on this dataset. To assess the effectiveness of our proposal under the condition of limited training data availability, we undersampled the data by taking one axial slice from consecutive slices in 15-20 mm intervals in the original dataset, such that similar numbers of slices were taken from each patient. The sampled dataset contains 1095 CT images of 20 subjects.

\subsection{Evaluation Metrics} 

The following metrics were calculated for each test image separately and then averaged. The first group were segmentation metrics: pixel-level precision, recall, and f-score. Besides, to evaluate the shape and
topological correctness of segmented objects, two types of
topological metrics were calculated on the predicted objects the
majority part of which overlapped with the ground truth: Betti number errors $\beta^\text{err}_0$ and $\beta^\text{err}_1$, respectively calculated for 0- and 1-dimensional homology, and their sum $\beta^\text{err}$; and Betti matching errors $\mu^\text{err}_0$ and $\mu^\text{err}_1$, respectively calculated for 0- and 1-dimensional homology, and their sum $\mu^\text{err}$~\cite{stucki2023topologicallyfaithfula}. Note that all topological metrics were evaluated on the contours of the ground truth and predictions. 

\subsection{Comparison Algorithms} 

Seven methods were used. The first was the baseline that used the cross entropy loss, ${\rm CE}$, in training. \textit{ActiveContourLoss}~\cite{chen2019learning} and \textit{HausdorffDistanceLoss}~\cite{karimi2019reducing} presented shape preserving losses. We chose them to understand the importance of preserving objects' topology, which can simultaneously model shape and geometry, instead of emphasizing only the shape. \textit{TopologyLossLikelihoodFiltration}~\cite{hu2019topologypreservingdeep} and~\textit{TopologyLossDistanceFiltration}~\cite{ozcelik2023topologyawareloss} also defined topology preserving losses. However, they measured topological dissimilarity by the Wasserstein distance calculated on the persistence diagrams of ground truth and prediction maps; they employed neither the persistence image representation nor the adaptive scheduler mechanism. We used them to understand the advantageous of using the proposed adaptive persistence image representation over the direct use of persistence diagrams and the Wasserstein distance. Another topology-preserving method, \textit{clDiceTopologyPreservingLoss}~\cite{shit2021cldicenovel}, was also included to understand the true capability of persistent homology over topological correctness. This method paid attention to the correctness of the morphological skeletons of objects with the topological precision and recall terms that were generated by soft-skeletonization found in the clDice loss. \textit{NonAdaptivePersistenceImageLoss} was for an ablation study to show the importance of the proposed adaptive scheduler mechanism; it used the proposed persistence image loss but did not adjust the weighting function adaptively based on the network's performance.

\subsection{Network Architecture and Implementation Details} 

For our model and all comparison methods, we used the same TransUnet architecture~\cite{chen2021transunettransformers}, a well-known architecture with transformer layers. To eliminate possible divergence in their training, a warm-up period was used for all methods, where the network was trained only using cross entropy for the first 10 epochs. The hyperparameters of each method were separately tuned on the validation set using the weights and biases module~\cite{wandb}; the test fold was not used in this process. For all methods, we used three-fold cross-validation and got five runs, starting with different initial network weights, for each fold. Since weight initialization affects the final network, we used the same set of seeds in random number generation for fair comparisons. We used PyTorch for network implementations and the Gudhi module~\cite{gudhi} for topological operations.

\begin{table*}[!ht]
\centering
\caption{For the in-house dataset, the test fold metrics obtained when TransUnet is used. Results are the averages of the total 15 runs from 3 folds and their standard deviations. $\uparrow$ and $\downarrow$ signs indicate performance metrics that ``bigger is better" and ``smaller is better", respectively. Statistically significantly best metrics ($p=0.05$) are indicated in bold. Note that for a selected metric, there is no statistically significant difference between the values that are all shown in bold.}
\vspace{-0.1cm}
\label{table:TransUnet_results}
\renewcommand{\arraystretch}{1.1} 
\begin{tabular}{lc@{\hspace{2.2mm}}c@{\hspace{2.2mm}}cc@{\hspace{2.2mm}}c@{\hspace{2.2mm}}c@{\hspace{2.2mm}}c@{\hspace{2.2mm}}c@{\hspace{2.2mm}}c}
\toprule

& \multicolumn{3}{c}{Segmentation Metrics} & \multicolumn{6}{c}{Topological Metrics} \\ 
\cmidrule(lr){2-4} \cmidrule(l){5-10}

Method  & Precision $\uparrow$ & Recall $\uparrow$ & F-score $\uparrow$ & $\beta^\text{err}_0$ $\downarrow$ & $\beta^\text{err}_1$ $\downarrow$ & $\beta^\text{err}$ $\downarrow$ & $\mu^\text{err}_0$ $\downarrow$ & $\mu^\text{err}_1$ $\downarrow$ & $\mu^\text{err}$ $\downarrow$ \\ \midrule

Baseline  & \textbf{86.25$\pm$2.56} & 78.43$\pm$2.98 & 80.09$\pm$1.49 & 0.80$\pm$0.09 & 0.78$\pm$0.09 & 1.58$\pm$0.17 & 0.84$\pm$0.10 & 1.19$\pm$0.16 & 2.03$\pm$0.24 \\  
ActiveContourLoss~\cite{chen2019learning}  & \textbf{86.55$\pm$2.12} & 79.36$\pm$4.53 & 80.71$\pm$2.04 & 0.79$\pm$0.12 & 0.76$\pm$0.12 & 1.55$\pm$0.24 & 0.82$\pm$0.13 & 1.17$\pm$0.16 & 1.99$\pm$0.28 \\
HausdorffDistanceLoss~\cite{karimi2019reducing} & 85.30$\pm$2.39 & 80.11$\pm$2.63 & 80.64$\pm$1.38 & 0.73$\pm$0.09 & 0.72$\pm$0.08 & 1.45$\pm$0.17 & 0.77$\pm$0.09 & \textbf{1.09$\pm$0.13} & 1.86$\pm$0.21 \\
TopologicalLossLikelihoodFilt~\cite{hu2019topologypreservingdeep}  & 84.15$\pm$2.50 & 80.71$\pm$2.66 & 80.28$\pm$1.61 & 0.74$\pm$0.09 & 0.73$\pm$0.08 & 1.47$\pm$0.17 & 0.78$\pm$0.09 & 1.15$\pm$0.08 & 1.93$\pm$0.16 \\
TopologicalLossDistanceFilt~\cite{ozcelik2023topologyawareloss} & 84.35$\pm$3.28 & 80.42$\pm$2.65 & 80.31$\pm$1.94 & 0.74$\pm$0.08 & 0.72$\pm$0.09 & 1.46$\pm$0.17 & 0.78$\pm$0.08 & 1.11$\pm$0.12 & 1.89$\pm$0.20 \\
clDiceTopologyPreservingLoss~\cite{shit2021cldicenovel} & 84.00$\pm$2.35 & 81.11$\pm$3.87 & 80.59$\pm$1.59 & 1.92$\pm$0.59 & \textbf{0.63$\pm$0.10} & 2.55$\pm$0.60 & 2.08$\pm$0.63 & \textbf{1.09$\pm$0.18} & 3.17$\pm$0.60 \\
NonAdaptivePersistenceImageLoss  & 85.61$\pm$2.69 & 80.11$\pm$2.91 & 80.62$\pm$1.22 & 0.73$\pm$0.08 & 0.73$\pm$0.08 & 1.46$\pm$0.16 & 0.78$\pm$0.09 & 1.12$\pm$0.09 & 1.90$\pm$0.16 \\
\piatt:PersistenceImageLoss  & 85.25$\pm$2.17 & \textbf{82.54$\pm$1.71} & \textbf{82.10$\pm$0.87} & \textbf{0.67$\pm$0.06} & \textbf{0.65$\pm$0.05} & \textbf{1.32$\pm$0.11} & \textbf{0.70$\pm$0.06} & \textbf{1.06$\pm$0.08} & \textbf{1.76$\pm$0.12}   \\ \bottomrule
\end{tabular}
\end{table*}

\begin{table*}[t]
\centering
\vspace{0.15cm}

\caption{Test fold metrics when Unet is used on the in-house dataset and when TransUnet is used on the AVT-Kits dataset. Statistically significantly best metrics ($p < 0.05$) are indicated in bold.}
\label{table:ablation}
\vspace{-0.15cm}
\begin{tabular}{lc@{\hspace{2.5mm}}c@{\hspace{2.5mm}}cc@{\hspace{2.5mm}}c@{\hspace{2.5mm}}c}
\toprule
& \multicolumn{3}{c}{Unet on } & \multicolumn{3}{c}{TransUnet on} \\ 
& \multicolumn{3}{c}{in-house dataset} & \multicolumn{3}{c}{AVT-Kits dataset} \\ 
\cmidrule(lr){2-4} \cmidrule(l){5-7}
Methods  & F-score  $\uparrow$  & $\beta^\text{err}$ $\downarrow$  & $\mu^\text{err}$ $\downarrow$  & F-score  $\uparrow$  & $\beta^\text{err}$ $\downarrow$  & $\mu^\text{err}$ $\downarrow$ \\ \midrule
Baseline  & 77.04$\pm$3.22 & 1.86$\pm$0.33 & 2.38$\pm$0.36 & 79.92$\pm$1.75 & 0.92$\pm$0.10 & 1.18$\pm$0.12 \\
ActiveContourLoss~\cite{chen2019learning} & 77.78$\pm$2.76 & 1.93$\pm$0.38 & 2.44$\pm$0.42 & 79.64$\pm$1.85 & 0.97$\pm$0.09 & 1.22$\pm$0.11 \\
HausdorffDistanceLoss~\cite{karimi2019reducing} & 77.28$\pm$3.36 & 1.89$\pm$0.34 & 2.47$\pm$0.39 & 79.04$\pm$2.14 & 0.96$\pm$0.10 & 1.23$\pm$0.13 \\
TopologicalLossLikelihoodFilt~\cite{hu2019topologypreservingdeep} & 77.84$\pm$3.11 & 1.82$\pm$0.29 & 2.39$\pm$0.37 & 79.77$\pm$1.50 & 0.95$\pm$0.09 & 1.21$\pm$0.11 \\
TopologicalLossDistanceFilt~\cite{ozcelik2023topologyawareloss} & 77.26$\pm$2.73 & 1.80$\pm$0.30 & 2.35$\pm$0.33 & 79.47$\pm$1.55 & 0.95$\pm$0.10 & 1.21$\pm$0.11 \\
clDiceTopologyPreservingLoss~\cite{shit2021cldicenovel} & 79.52$\pm$1.90 & 1.56$\pm$0.15 & 2.08$\pm$0.21 & 79.50$\pm$2.28 & 0.91$\pm$0.10 & 1.22$\pm$0.14 \\
NonAdaptivePersistenceImageLoss & 77.38$\pm$3.10 & 1.78$\pm$0.33 & 2.28$\pm$0.31 & 80.13$\pm$1.43 & 0.93$\pm$0.11 & 1.18$\pm$0.14 \\
\piatt:PersistenceImageLoss & \textbf{80.47$\pm$1.02} & \textbf{1.44$\pm$0.18} & \textbf{1.90$\pm$0.22} & \textbf{81.19$\pm$0.86} & \textbf{0.82$\pm$0.06} & \textbf{1.05$\pm$0.08} \\ \bottomrule

\end{tabular}
\vspace{-0.35cm}
\end{table*}

%% file: parts/5_results_discussion.tex
\section{Results and Discussion}

For the in-house dataset, Table~\ref{table:TransUnet_results} reports the 
average and standard deviation of the test fold metrics obtained for the 15 runs (three folds, and five runs for each fold). This table reveals that all methods surpassed the baseline, indicating the benefit of incorporating shape or topology awareness into a loss function. On the other hand, our proposed loss gave the best f-score, Betti number errors, and Betti matching errors. These results were statistically significant with $p < 0.05$ when the paired-sample t-test was applied. Moreover, it led to low standard deviations, indicating stability in network training. 

Compared with \textit{ActiveContourLoss} and \textit{HausdorffDistanceLoss}, which only emphasized shape awareness, the improvement suggested the usefulness of simultaneously modeling shape and geometry through topology. \textit{TopologicalLossLikelihoodFiltration} and \textit{TopologicalLossDistanceFiltration} also enforced the network to minimize topological dissimilarity between the ground truth and prediction maps. However, they achieved this by calculating their persistence diagrams and measuring the Wasserstein distance between them. Worse f-scores and Betti number/matching errors might be attributed to the problems associated with matching the points in the two diagrams. The persistence image representation eliminated the requirement of point matching and mitigated the problem. \textit{clDiceTopologyPreservingLoss} preserved topological correctness by integrating the intersection of objects' skeletons into the loss. Table~\ref{table:TransUnet_results} shows that this skeletonization technique improved the topological metrics for 1-dimensional homology. Nevertheless, for these metrics, there were no statistically significant difference between \textit{clDiceTopologyPreservingLoss} and our model. Moreover, our model gave statistically significantly better results for the other performance metrics.

\begin{figure*}[t]
    \begin{center}
    \begin{tabular}{@{~}c@{~}c@{~}c@{~}c@{~}c@{~}c@{~}c@{~}c@{~}c@{~}}
    \includegraphics[width =0.193\columnwidth]{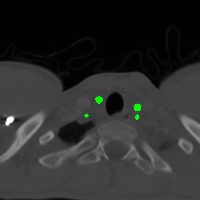} &
    \includegraphics[width =0.193\columnwidth]{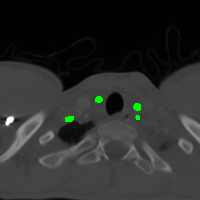} &
    \includegraphics[width =0.193\columnwidth]{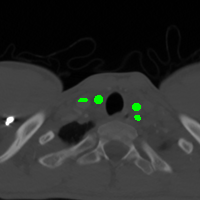} &
    \includegraphics[width =0.193\columnwidth]{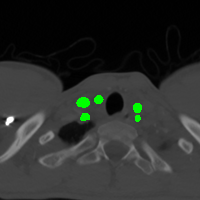} &
    \includegraphics[width =0.193\columnwidth]{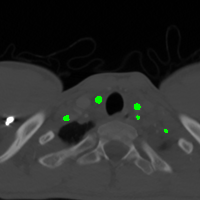} &
    \includegraphics[width =0.193\columnwidth]{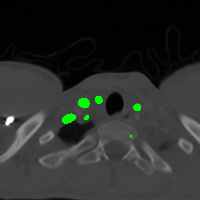} &
    \includegraphics[width =0.193\columnwidth]
    {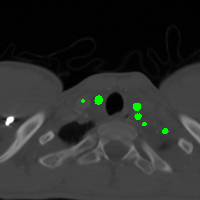} &
    \includegraphics[width =0.193\columnwidth]{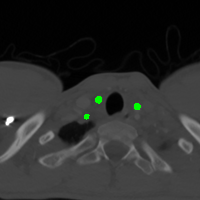} &
    \includegraphics[width =0.193\columnwidth]{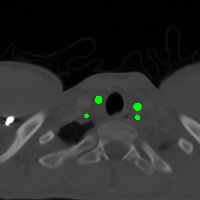} 
    \vspace{-0.03cm} \\
    \includegraphics[width =0.193\columnwidth]{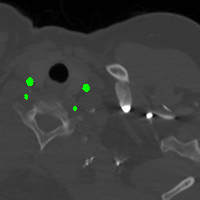} &
    \includegraphics[width =0.193\columnwidth]{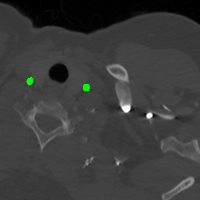} &
    \includegraphics[width =0.193\columnwidth]{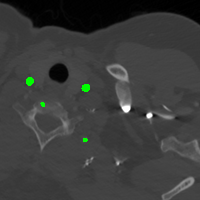} &
    \includegraphics[width =0.193\columnwidth]{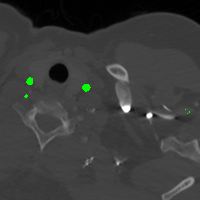} &
    \includegraphics[width =0.193\columnwidth]{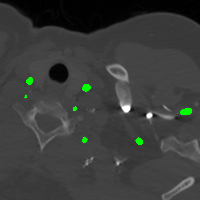} &
    \includegraphics[width =0.193\columnwidth]{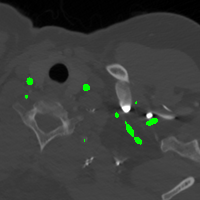} &
    \includegraphics[width =0.193\columnwidth]
    {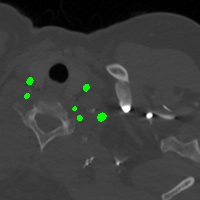} &
    \includegraphics[width =0.193\columnwidth]{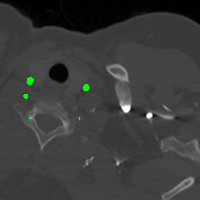} &
    \includegraphics[width =0.193\columnwidth]{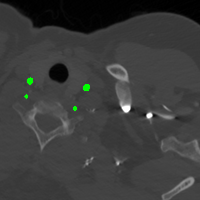} 
    \vspace{-0.03cm} \\
    \includegraphics[width =0.193\columnwidth]{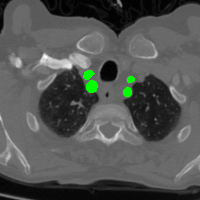} &
    \includegraphics[width =0.193\columnwidth]{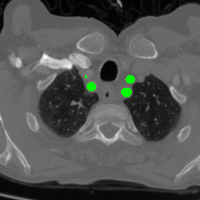} &
    \includegraphics[width =0.193\columnwidth]{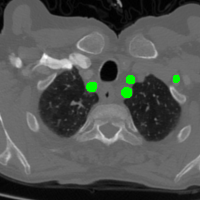} &
    \includegraphics[width =0.193\columnwidth]{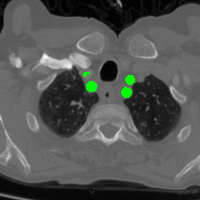} &
    \includegraphics[width =0.193\columnwidth]{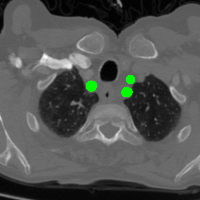} &
    \includegraphics[width =0.193\columnwidth]{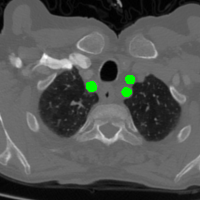} &
    \includegraphics[width =0.193\columnwidth]
    {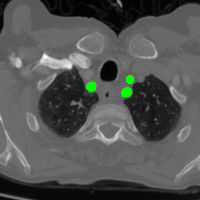} &
    \includegraphics[width =0.193\columnwidth]{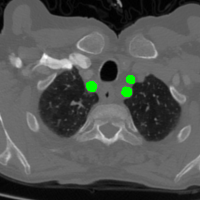} &
    \includegraphics[width =0.193\columnwidth]{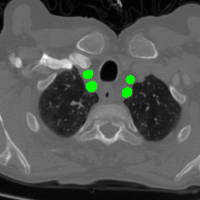} 
    \vspace{-0.03cm} \\
    \includegraphics[width =0.193\columnwidth]{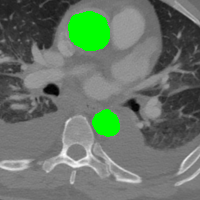} &
    \includegraphics[width =0.193\columnwidth]{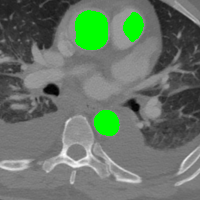} &
    \includegraphics[width =0.193\columnwidth]{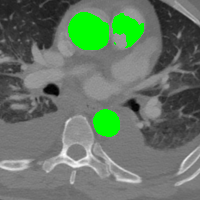} &
    \includegraphics[width =0.193\columnwidth]{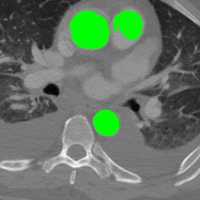} &
    \includegraphics[width =0.193\columnwidth]{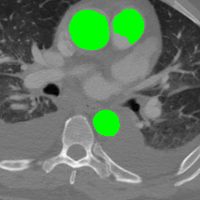} &
    \includegraphics[width =0.193\columnwidth]{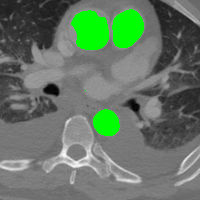} &
    \includegraphics[width =0.193\columnwidth]
    {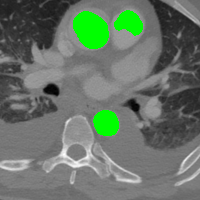} &
    \includegraphics[width =0.193\columnwidth]{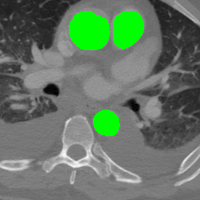} &
    \includegraphics[width =0.193\columnwidth]{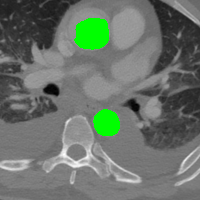} 
    \vspace{-0.03cm} \\
    \includegraphics[width =0.193\columnwidth]{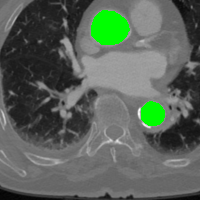} &
    \includegraphics[width =0.193\columnwidth]{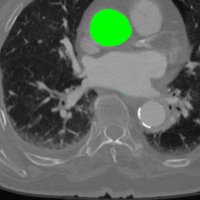} &
    \includegraphics[width =0.193\columnwidth]{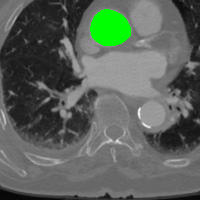} &
    \includegraphics[width =0.193\columnwidth]{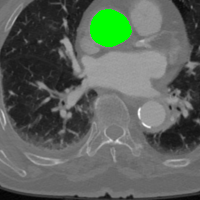} &
    \includegraphics[width =0.193\columnwidth]{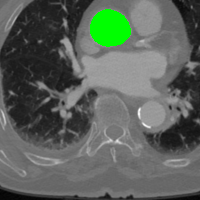} &
    \includegraphics[width =0.193\columnwidth]{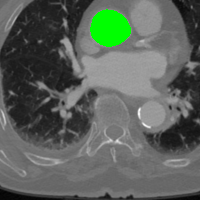} &
    \includegraphics[width =0.193\columnwidth]
    {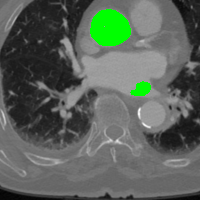} &
    \includegraphics[width =0.193\columnwidth]{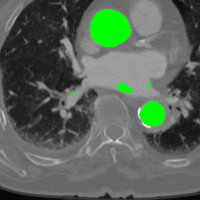} &
    \includegraphics[width =0.193\columnwidth]{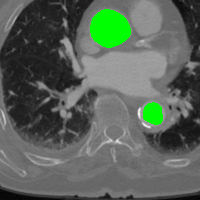} 
    \vspace{-0.03cm} \\
    (a) & (b) & (c) & (d) & (e) & (f) & (g) & (h) & (j) \vspace{-0.15cm}
    \end{tabular}
    \end{center}
    \vspace{-0.1cm}
    \captionsetup{font = small}
    \caption{For the in-house dataset, visual results on exemplary CT images. (a) Ground truths. Visual results obtained by (b) Baseline, (c) ActiveContourLoss~\cite{chen2019learning}, (d) HausdorffDistanceLoss~\cite{karimi2019reducing}, (e) TopologicalLossLikelihoodFiltration~\cite{hu2019topologypreservingdeep}, (f) TopologicalLossDistanceFiltration~\cite{ozcelik2023topologyawareloss}, (g) clDiceTopologyPreservingLoss~\cite{shit2021cldicenovel}, (h) NonAdaptivePersistenceImageLoss, and (j) proposed \piatt:PersistenceImageLoss. Vessel pixels are shown in green on the CT images.}
    \label{fig:TransUnet_results}
\end{figure*}

Although \textit{NonAdaptivePersistenceImageLoss} used the persistence image representation, it did not significantly improve the results of the other methods. One reason might be as follows: CT images contain larger aorta and relatively smaller great vessels. Consequently, holes associated with the aorta are expected to persist longer than those associated with the great vessels. Using a fixed weighting function may not adequately capture both characteristics at the same time. On the other hand, thanks to its adaptive scheduler mechanism, the proposed persistence image representation addressed this discrepancy successfully. It prioritized holes with longer lives in the first epochs, and gradually shifted its focus to those with shorter lives towards the end of training. This enhanced the model's ability to learn the overall geometry of all vessels, as also seen in the first three images of the visual results given in Fig.~\ref{fig:TransUnet_results}. Additionally, it helped our model predict the positions of more challenging objects, such as very small vessels (second and third images of Fig.~\ref{fig:TransUnet_results}) and confusing parts in inputs (upper right false object of the fourth image), which may not be captured by the other methods. This decreased false negative vessel predictions of our model, leading to an improvement in its recall and f-score metrics as well as in its Betti number/matching errors, especially for the 0-dimensional homology that generally corresponded to small and hard-to-find connected components. Moreover, our model yielded accurate results in predicting larger aorta objects, as demonstrated in the last image of Fig.~\ref{fig:TransUnet_results}.

\subsection{Another Network Architecture}
To better understand the generalizability and contribution of the \piatt~loss, we also repeated the experiments on a simpler network architecture, Unet~\cite{ronneberger2015unetconvolutional}. We used its original implementation except that we added dropout layers into its encoder. The quantitative metrics obtained over 15 runs are reported in Table~\ref{table:ablation}. (Due to page limits, only the f-score and overall Betti number/matching errors are reported, and no visual results are provided.) Here we observe that most of the comparison methods only slightly improved the baseline, unlike the case where they used TransUnet. The reason might be as follows: Due to the less complex architecture of Unet, the customized losses proposed by these methods may not always help encode shape and topology awareness, which may guide the network to converge a better solution. This was also observed in the standard deviations of the f-scores in Table~\ref{table:ablation}; for the comparison methods, they were much larger than the ones when TransUnet was used. On the other hand, thanks to the adaptive scheduler mechanism, the proposed \piatt~ loss led to a statistically significant increase ($p < 0.05$) in all performance metrics, regardless of the network complexity.

\begin{figure}
  \centering
  \vspace{-0.4cm}
\includegraphics[width=8.2cm]{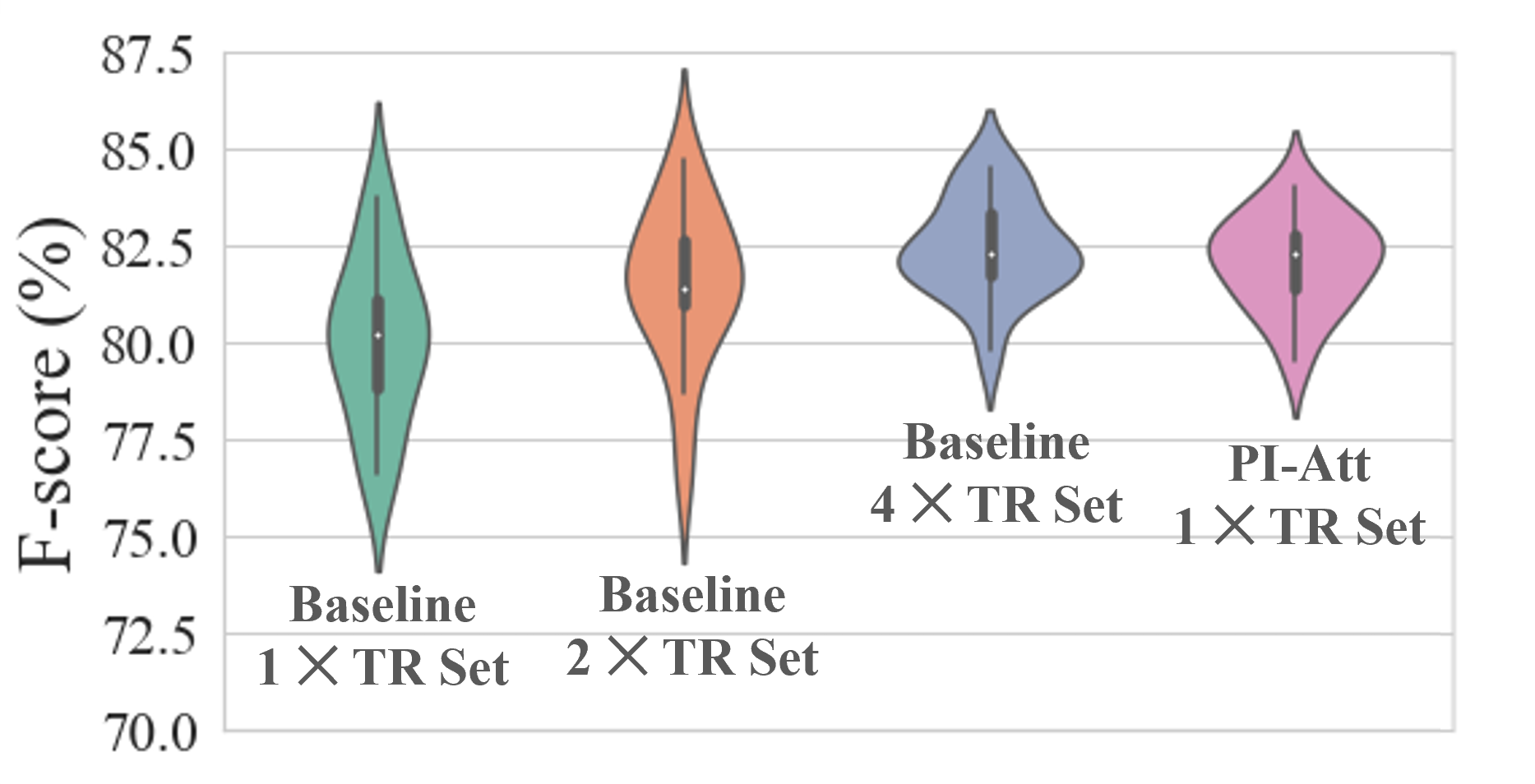} 
\captionsetup{font = small} 
\caption{Violin plots of the test fold f-scores for the in-house dataset when TransUnet is used. The first three are for the baseline trained on the original, doubled, and quadrupled training set, respectively. The last is for the proposed model trained on the original training set. They were generated for the 15 runs.} 
\label{fig:abl_data_size}
\vspace{-0.3cm} 
\end{figure}

\begin{figure}[ht]
    \centering
    \vspace{-0.3cm}    
    \begin{tabular}    {@{\hspace{0.3mm}}c@{\hspace{0.3mm}}c@{\hspace{0.3mm}}c@{\hspace{0.3mm}}c@{\hspace{0.3mm}}}
    \includegraphics[width = 0.190\columnwidth]{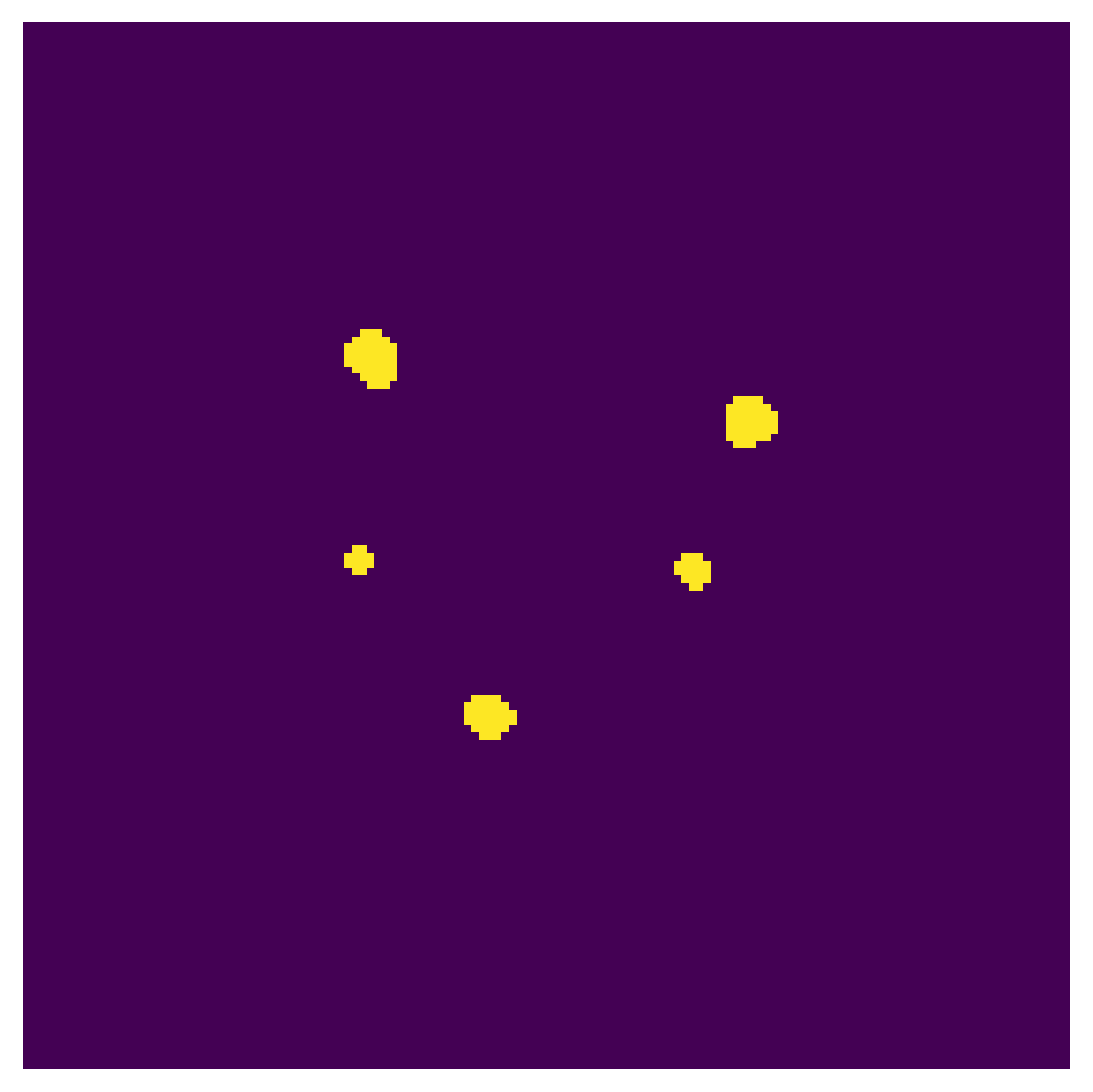} &
    \includegraphics[width = 0.190\columnwidth]{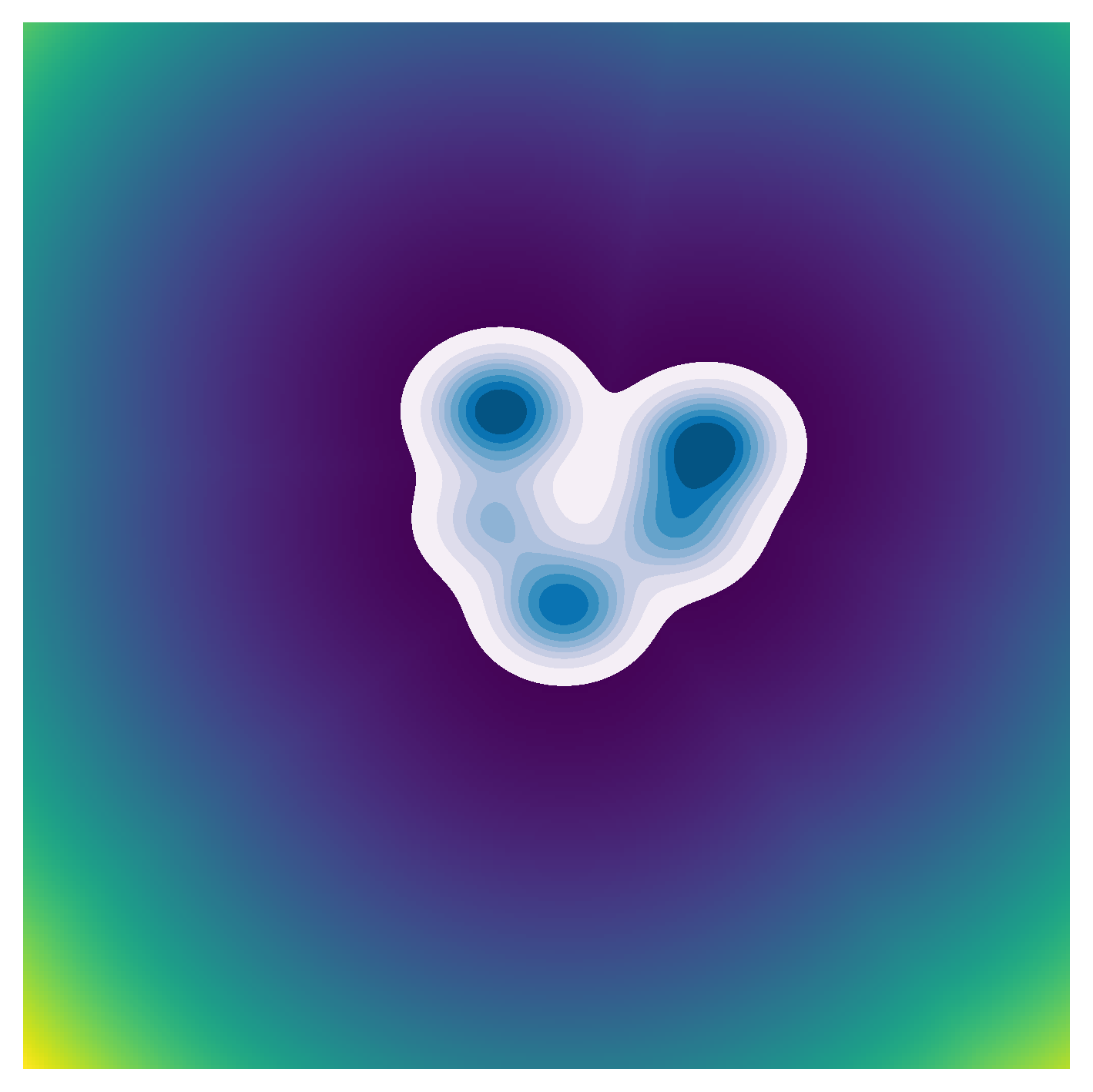} &
    \includegraphics[width = 0.190\columnwidth]{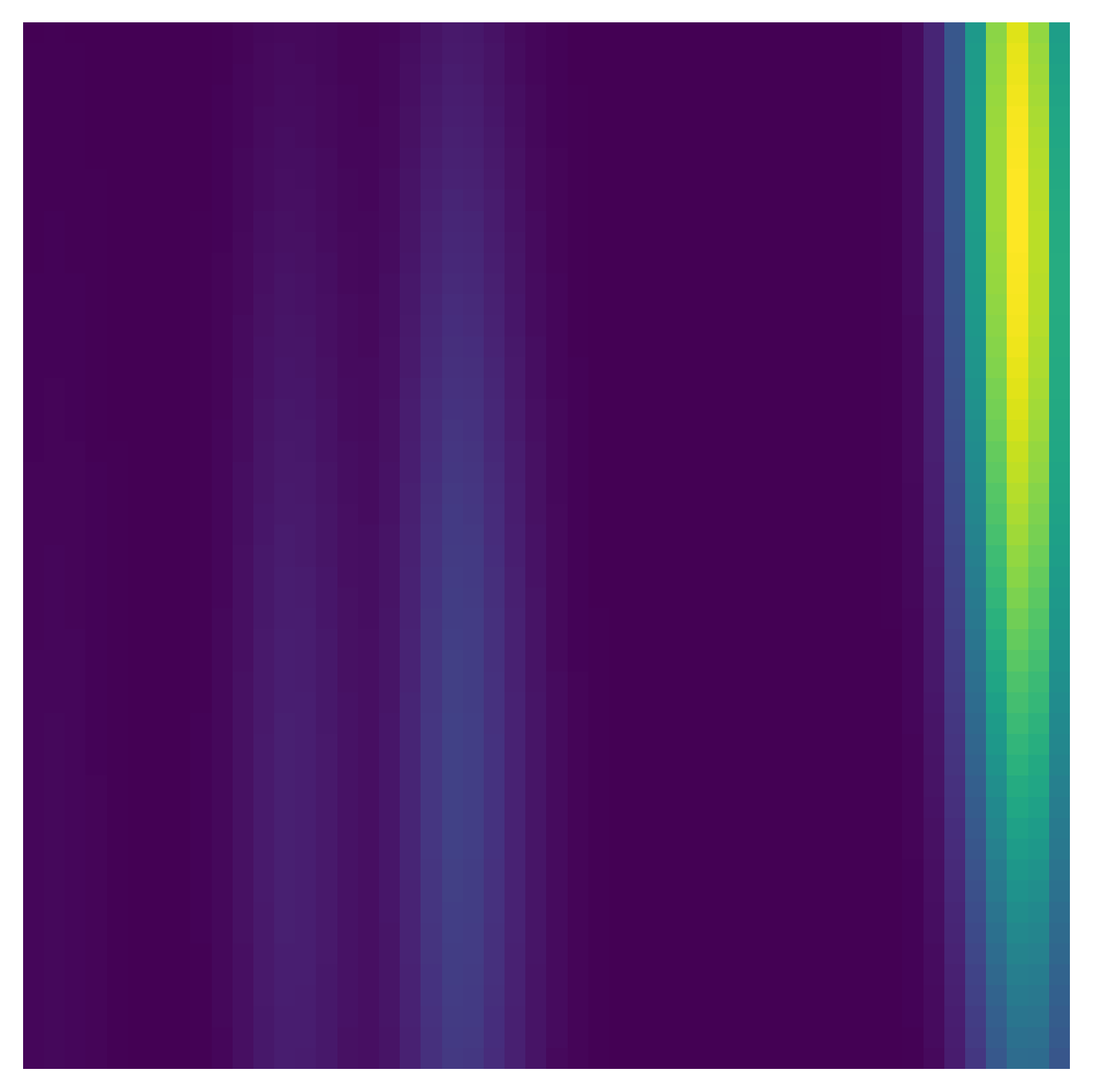} &
    \includegraphics[width = 0.190\columnwidth]{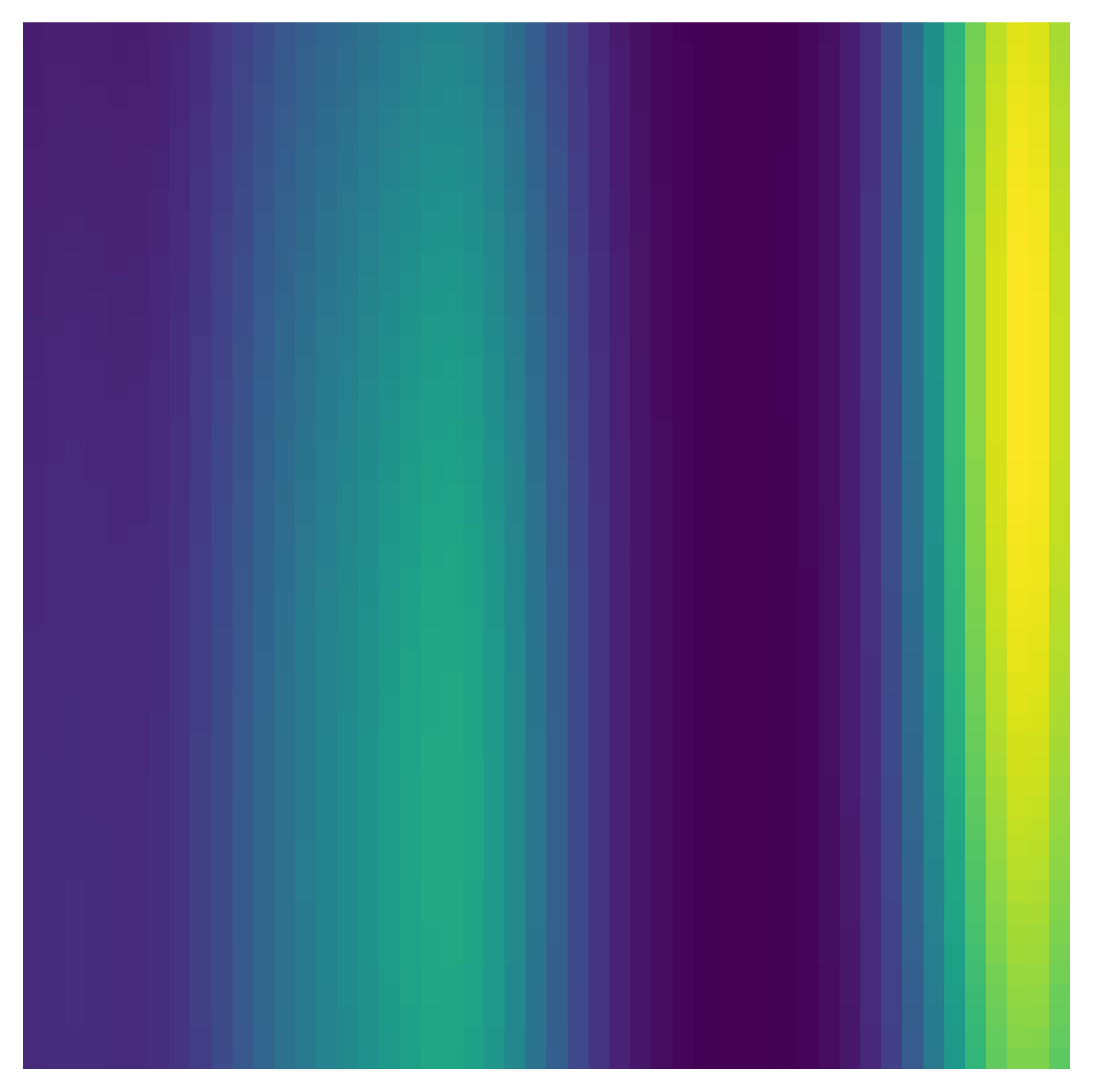} 
    \vspace{-0.09cm} \\ 
    &
    \includegraphics[width = 0.190\columnwidth]{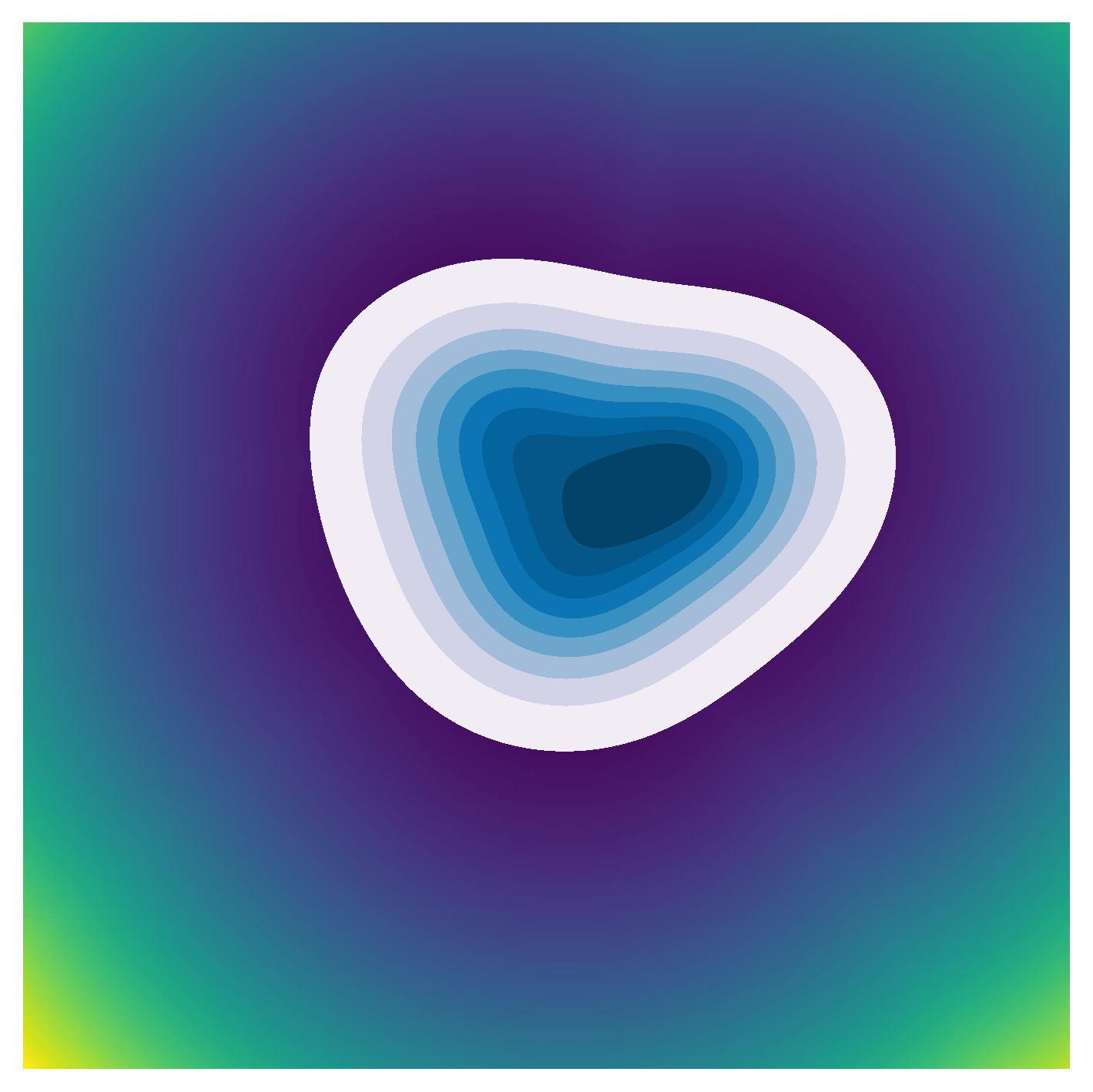} &
    \includegraphics[width = 0.190\columnwidth]{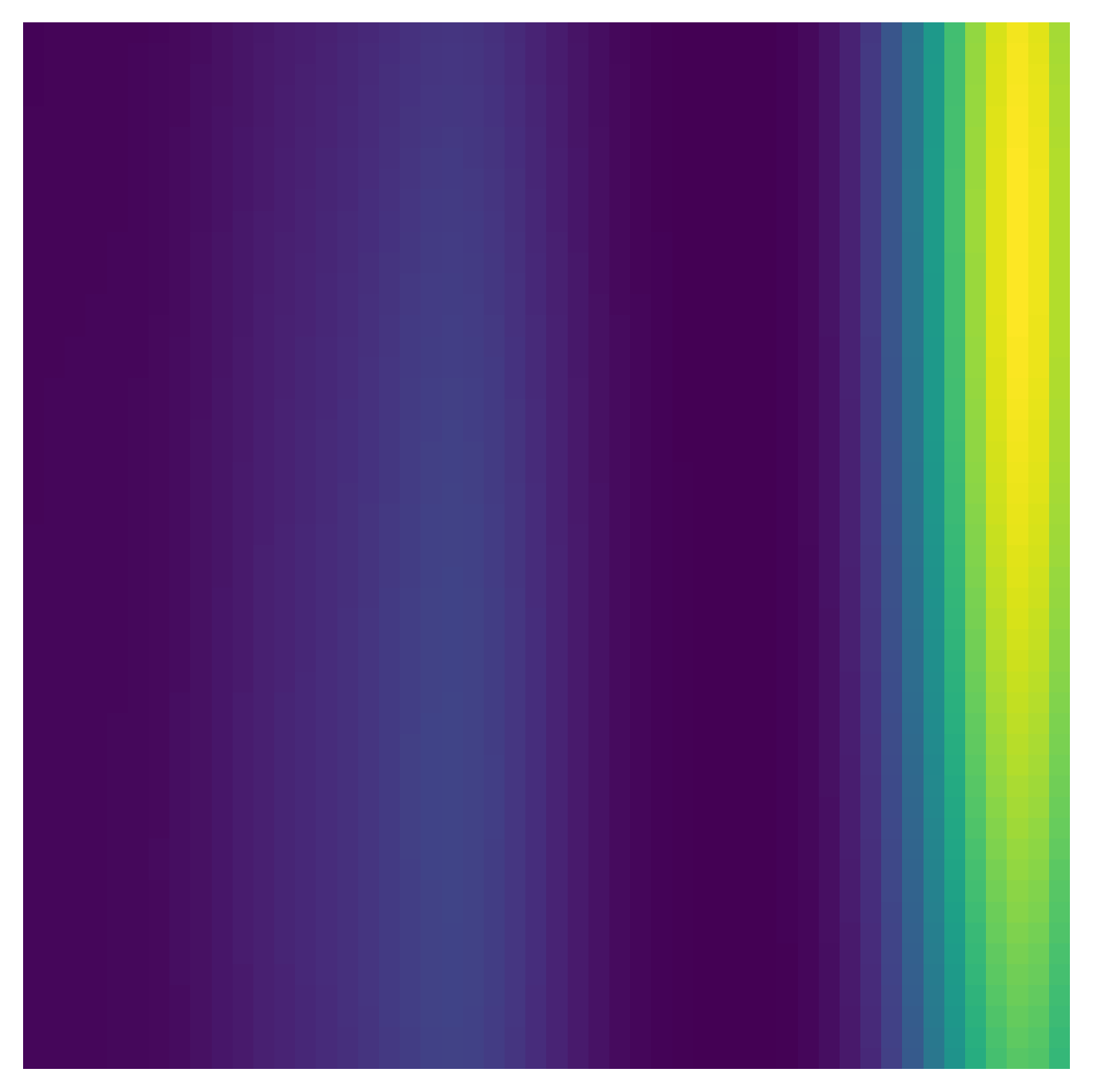} &
    \includegraphics[width = 0.190\columnwidth]{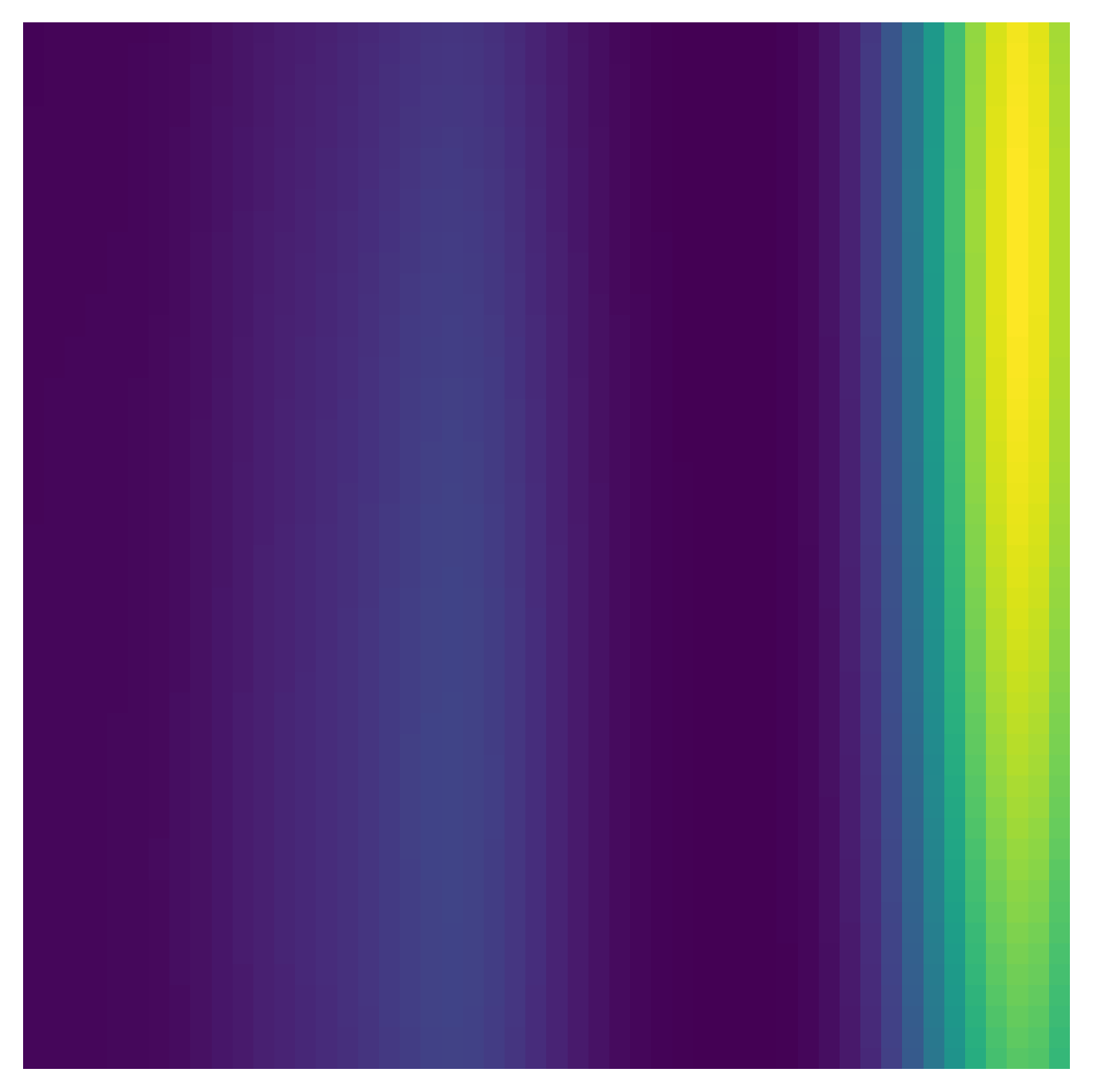} 
    \vspace{-0.09cm} \\  
    &
    \includegraphics[width = 0.190\columnwidth]{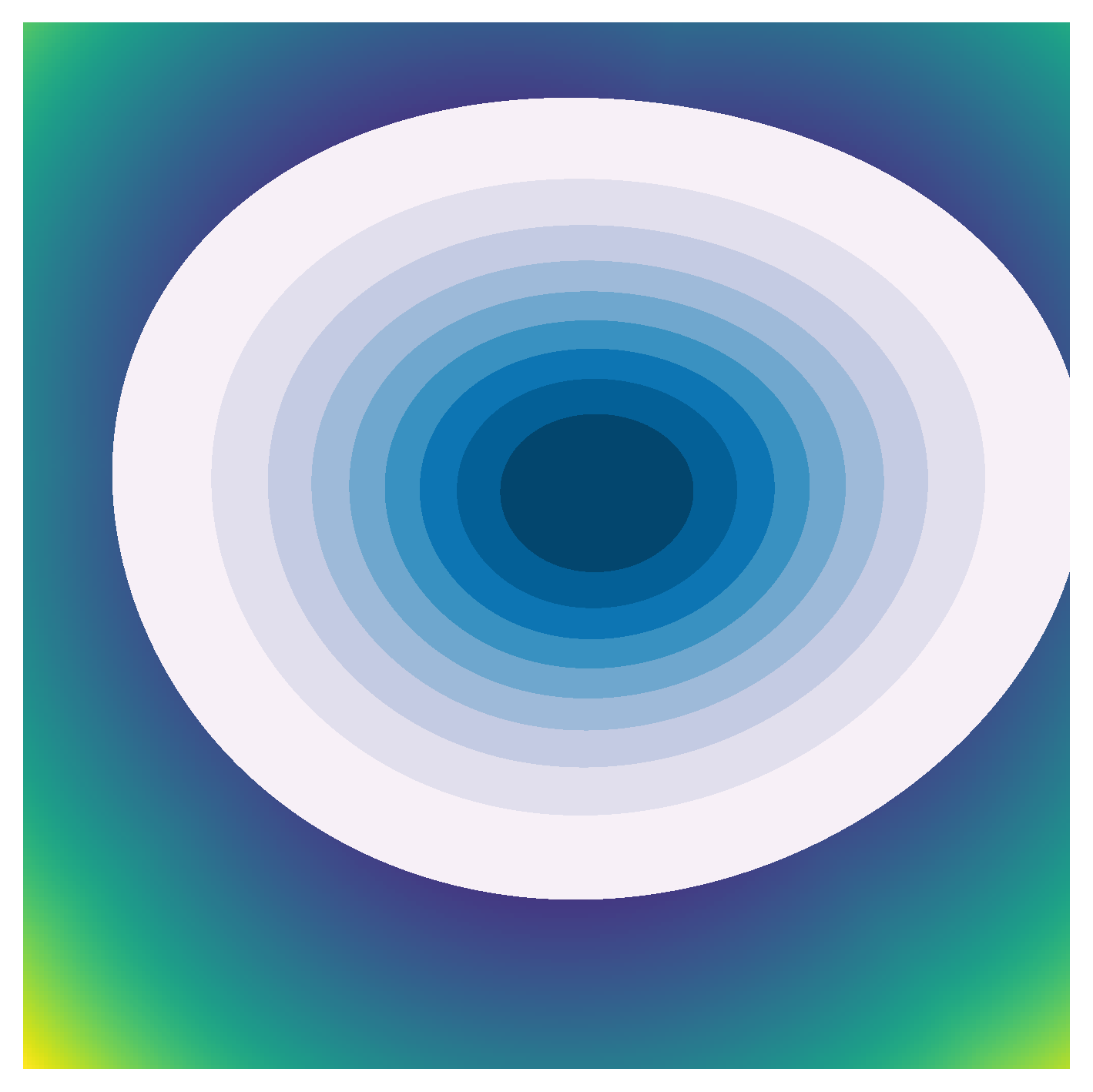} &
    \includegraphics[width = 0.190\columnwidth]{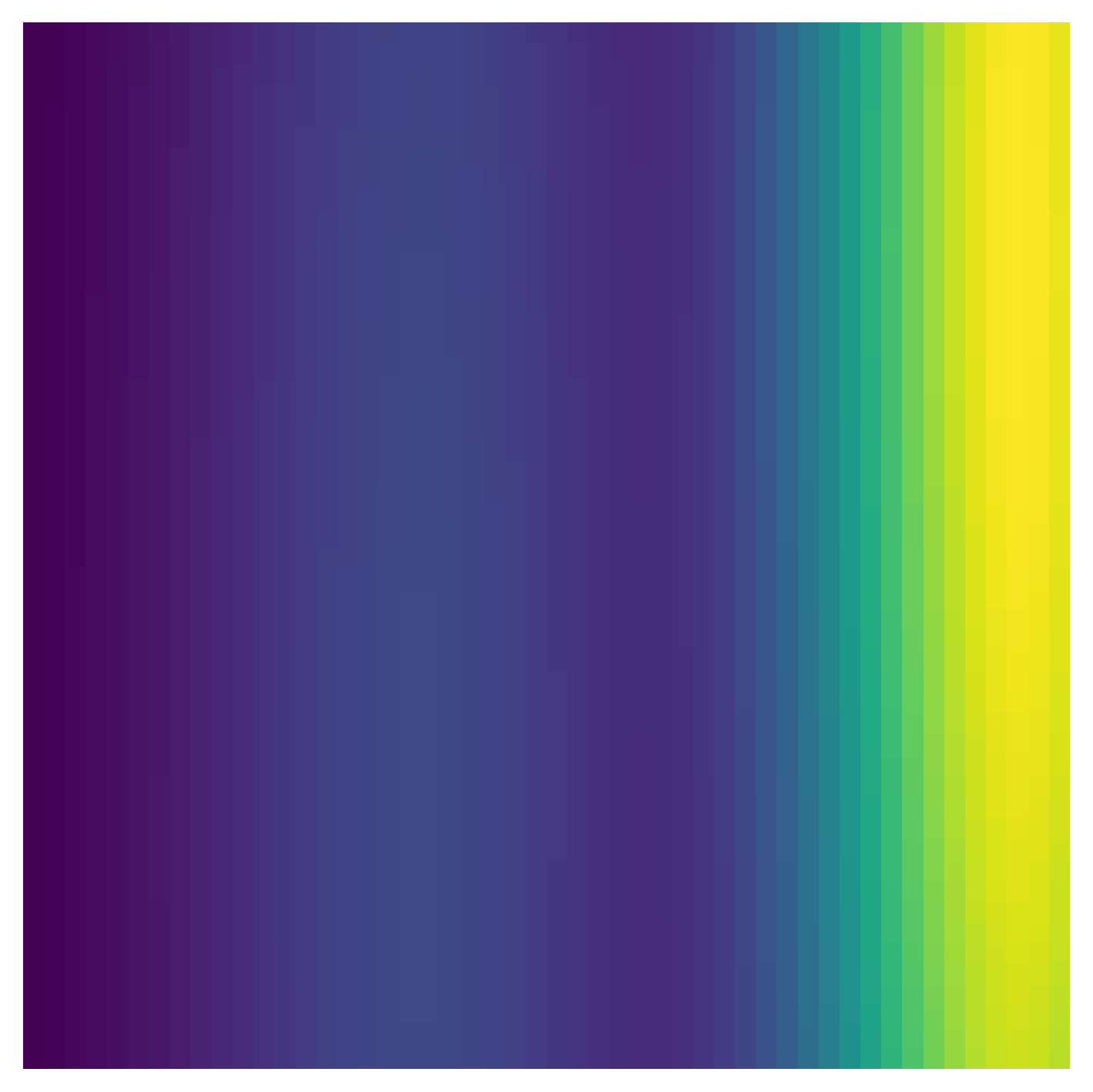} &
    \includegraphics[width = 0.190\columnwidth]{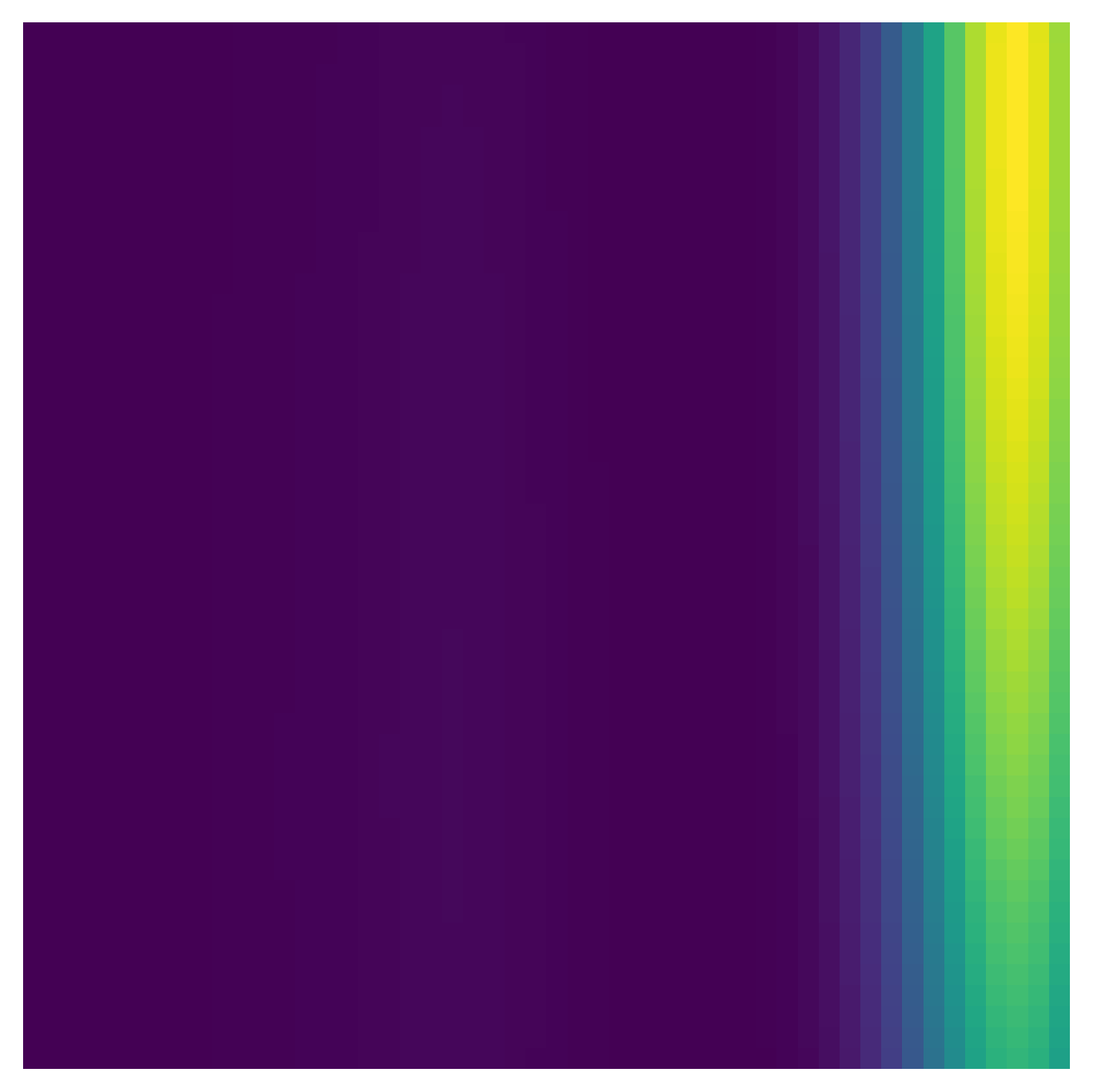} 
    \vspace{-0.01cm} \\ 
     (a) & (b) & (c) & (d)
    \end{tabular}
    \captionsetup{font = small}
    \caption{(a) An example map. Persistence images obtained with different (b) bandwidths $B$, (c) variances $\sigma^2$, and (d) weighting function parameters $\gamma^0$. The values of all these three parameters increase from top to bottom.} 
    \label{fig:persistence_image_param}
\vspace{-0.3cm}
\end{figure}

\begin{figure*}
\vspace{-0.1cm}
\centering
\includegraphics[width=18cm]{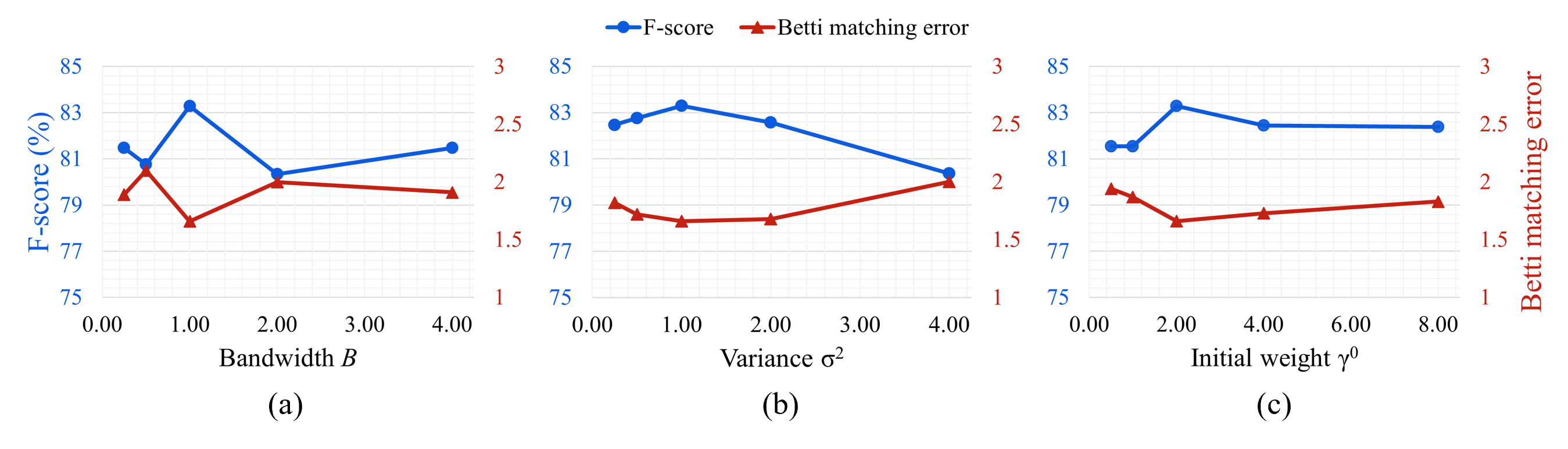} 
\vspace{-0.3cm}
\captionsetup{font = small}
\caption{Performance metrics as a function of the external parameters: (a) bandwidth $B$ of the density kernel estimator, (b) variance $\sigma^2$ of the Gaussian kernel, and (c) weighting function parameter $\gamma^0$. They were the average test results for the first fold when TransUnet was used for the in-house dataset.} 
\label{fig:abl_persistence_image}
\vspace{-0.3cm}
\end{figure*}

\subsection{Training with Limited Data vs. More Data} 

In medical image analysis domain, it is very typical to have limited annotated data due to the expert-required annotation process. We also explored the contribution of the \piatt~loss from this perspective. Previously, we compared our model with the baseline trained on the same amount of data. Next, we increased the training set size only for the baseline, by adding new annotated CT images, and compared it with the proposed model trained on the original training set. Fig.~\ref{fig:abl_data_size} shows the violin plots of the f-scores obtained by the baseline trained on the original training set as well on the doubled and quadrupled training sets. These were the results of 15 runs obtained with TransUnet for the in-house dataset. These plots show the following: First, the f-score and stability (small variation) of the baseline were improved with an increase in the training set size, as expected. Second, the proposed model trained on the original training set led to similar performance and stability with the baseline trained on the quadrupled training set. This experiment suggests that the proposed loss may serve as a regularizer in network training, even with limited data.

\subsection{Experiments on a Publicly Available Dataset} 

We repeated the experiments on the publicly available AVT dataset examples taken from KiTS19 Challenge, also prepared for vessel segmentation. Table~\ref{table:ablation} reports the test fold results obtained with TransUnet. It shows that the \piatt~loss improved the baseline. However, the other methods, which also incorporated shape or topology awareness in training, did not appear to improve the baseline. To analyze this further, we scrutinized the metrics obtained for each fold individually. We observed that all methods considerably improved the baseline for the first two test folds; our model yielded further improvements. On the other hand, for the third fold, the training set included images from atypical patients, which may cause the network to converge to a bad solution. Despite these subjects, our proposed model effectively handled the images of these subjects and converged to a good solution thanks to the regularization through the proposed loss function. 

\subsection{Parameter Analysis of Persistence Image Calculation}
\label{sec:param-analysis}

There are three external parameters affecting persistence image calculation: bandwidth $B$ of the density kernel estimator, variance $\sigma^2$ of the Gaussian kernel, and weighting function parameter $\gamma^0$. For an example ground truth map, Fig.~\ref{fig:persistence_image_param} illustrates the effects of changing these parameters on the resulting persistence image. Fig.~\ref{fig:abl_persistence_image} shows quantitative analyses; for each parameter, the values of the others were fixed and metrics were measured as a function of this parameter.

The first step of this calculation is to obtain the persistence diagram on a cubical complex after smoothing contour points by a Gaussian kernel with the bandwidth $B$. A large bandwidth gives a smooth density distribution, and in our case leading to continuous filtration values but also losing details of the geometry between the objects (last row of Fig.~\ref{fig:persistence_image_param}(b)). On the contrary, a small bandwidth gives an unsmooth distribution, which enhances the details but may also cause noisy calculations due to the discrete nature of the contour pixels in a digital image. We obtained the best performance with $B = 1.0$, which was a good trade-off between these two. It is worth noting that as seen in Fig.~\ref{fig:abl_persistence_image}(a), $B$ values that could not maintain this trade-off but adequately addressed at least one phenomenon, either leading to truly continuous filtrations (largest $B$ in the plot) or emphasizing the shape details (smallest $B$ in the plot), gave better results than those that neither maintained the trade-off nor addressed 
one of two phenomena.

The next step is to map each point in the diagram to the persistence image based on its lifetime $y$ and the weighting function $\omega(y, \gamma)$ (Lines 6-7 of Algorithm~\ref{algo:pers-img}). The variance $\sigma^2$ of the Gaussian spreads the point to a symmetric window. Smaller $\sigma^2$ values result in diagram points being represented separately in the persistence image (separate vertical light blue regions seen in the first map of Fig.~\ref{fig:persistence_image_param}(c)). However, this also causes to separately represent noisy points, which may mislead topological difference calculation. On the other hand, larger values better handle these noisy pixels but this time may generate one merged region for true points closely located in the diagram. This yields crucial information loss. Both of them decreased the performance as seen in Fig.~\ref{fig:abl_persistence_image}(b). 

The $\gamma$ parameter and the lifetime $y$ of a point determine how intensely this point is represented in the persistence image (Eqn.~\ref{eqn:omega}). Our work starts with an initial weight $\gamma^0$ and changes it adaptively during training. One can give similar importance to every object in the first epochs by setting $\gamma^0 = 1$. Since this forces the network to learn details before learning the outline, it can lead to learning noisy pixels, which greatly reduces the performance. On the other hand, too large $\gamma^0$ values favor learning the outline too much, and the adaptive scheduler mechanism may not find time to reduce it enough before the network converges to integrate points with relatively short lifetimes, which correspond to the details. This also reduces performance. In our experiments, we set $\gamma^0 = 2.0$ to focus on the shape of bigger objects and the geometry outline of smaller and close ones in the first epochs. With the proposed adaptive update mechanism, we can adequately emphasized the details in the following epochs. Fig.~\ref{fig:abl_persistence_image}(c) shows that selecting this $\gamma^0$ value eliminated the negative effects of noisy diagram points. Although larger $\gamma^0$ values led to similar f-scores, the Betti matching scores became worse, which indicated worsening in the topological correctness (e.g., in the topological details such as small object shapes and close objects positions).

%% file: parts/6_conclusion.tex
\section{Conclusion}
We proposed the first-time incorporation of the persistence image representation in a segmentation loss, \piatt, to integrate topological awareness into a segmentation network. In the proposed scheme, the weighting function of the persistence image representation is altered during training, by introducing an adaptive scheduler mechanism, considering pixel-wise performance and topological correctness of the prediction. Calculating topological dissimilarity between the ground truth and prediction via their persistence images mitigates the problems of using the Wasserstein distance, which often arise from incorrect point matches between two persistence diagrams. Besides, the proposed scheduler mechanism enables the network to start with learning topology outline and focus on topology details in later epochs. This facilitates continuous and steady progress in performance throughout training. Working on the problem of aorta and great vessel segmentation in CT images, our experiments on two datasets and with two network architectures revealed that the proposed \piatt~loss improved the performance of the baseline and the comparison methods with shape/topology preserving losses. Besides, it yielded superior results compared with the methods that integrated topology awareness into network design, but in the form of directly using persistence diagrams and the Wasserstein distance. This indicated the effectiveness of the persistence image representation and the proposed adaptive scheduler mechanism, which is only made possible through the persistence image representation.